\documentclass[useAMS,usenatbib]{mn2e}
\usepackage{subfigure} 
\usepackage{graphicx}
\usepackage[english]{babel}
\usepackage{graphicx}
\usepackage{float}
\usepackage{floatflt}
\usepackage{rotating}
\usepackage{lscape}
\usepackage{epsfig}

\title[A tale of two GRB-SNe at a common redshift of $z=0.54$]{A tale of two GRB-SNe at a common redshift of $z = 0.54$\footnote{Based on observations made with the NASA/ESA Hubble Space Telescope, obtained at the Space Telescope Science Institute, which is operated by the Association of Universities for Research in Astronomy, Inc., under NASA contract NAS 5-26555. These observations are associated with program \# 10909.}}

\author[Z. Cano et al.]
{Z. Cano$^{1}$\thanks{E-mail:zec@astro.livjm.ac.uk}, D. Bersier$^1$, C. Guidorzi$^{2,1}$, R. Margutti$^{3}$, K.M Svensson$^4$, S. Kobayashi$^1$, 
\newauthor
A. Melandri$^{1,3}$, K. Wiersema$^5$, A. Pozanenko$^{6}$, A.J. van der Horst$^{7,8}$, G. G. Pooley$^{9}$, 
\newauthor
A. Fernandez-Soto$^{10}$, A.J. Castro-Tirado$^{11}$, A. de Ugarte Postigo$^{12}$,  M. Im$^{13}$,  
\newauthor
A.P. Kamble$^{14}$, D. Sahu$^{15}$, J. Alonso-Lorite$^{16}$, G. Anupama$^{15}$, J. L. Bibby$^{17,1}$, 
\newauthor
 M. J. Burgdorf$^{18,19}$, N. Clay$^1$, P.A. Curran$^{20}$, T. A. Fatkhullin$^{21}$, A. S. Fruchter$^{22}$, 
\newauthor
 P. Garnavich$^{23}$, A. Gomboc$^{24,25}$, J. Gorosabel$^{11}$, J. F. Graham$^{22}$, U. Gurugubelli$^{15}$, 
\newauthor
 J. Haislip$^{26}$, K. Huang$^{27}$, A. Huxor$^{28}$, M. Ibrahimov$^{29}$, Y. Jeon$^{13}$, Y-B. Jeon$^{30}$, 
\newauthor
 K. Ivarsen$^{26}$,  D. Kasen$^{31,32}$, E. Klunko$^{33}$, C. Kouveliotou$^{34}$, A. LaCluyze$^{26}$, A. J. Levan$^4$,  
\newauthor
V. Loznikov$^6$, P.A. Mazzali$^{35,36,37}$, A. S. Moskvitin$^{21}$, C. Mottram$^1$, C. G. Mundell$^1$,  
\newauthor
P.E. Nugent$^{38}$, M. Nysewander$^{22}$,  P. T. O'Brien$^5$, W. -K. Park$^{13}$, V. Peris$^{16}$,
\newauthor
E. Pian$^{36,39}$, D. Reichart$^{26}$, J. E. Rhoads$^{40}$,  E. Rol$^{14}$, V. Rumyantsev$^{41}$,  
\newauthor
V. Scowcroft$^{42}$,  D. Shakhovskoy$^{41}$,  E. Small$^1$,  R. J. Smith$^{1}$, V. V. Sokolov$^{21}$,  
\newauthor
R.L.C. Starling$^5$, I. Steele$^1$,  R. G. Strom$^{14,43,44}$,  N. R. Tanvir$^5$, Y. Tsapras$^{45,46}$, 
\newauthor
Y. Urata$^{47}$, O. Vaduvescu$^{48,49}$, A. Volnova$^{50}$,  A. Volvach$^{41}$, R. A. M. J. Wijers$^{14}$, 
\newauthor
S. E. Woosley$^{31}$, D. R. Young$^{51}$\\ 
\scriptsize \noindent $^1$Astrophysics Research Institute, Liverpool John Moores University, Liverpool, UK. 
\scriptsize$^2$Dipartimento di Fisica, Universit\' a di Ferrara, via Saragat 1, I-44100 Ferrara, Italy. \\
\scriptsize$^{3}$INAF - Osservatorio Astronomico di Brera, via E. Bianchi 46, 23807, Merate, LC, Italy. 
\scriptsize$^4$Department of Physics, University of Warwick, Coventry UK. \\
\scriptsize$^5$Department of Physics and Astronomy, University of Leicester, Leicester, UK. 
\scriptsize$^{6}$Space Research Institute of RAS, Profsoyuznaya, 84/32, Moscow, Russia. \\
\scriptsize$^7$NASA/Marshall Space Flight Center, Huntsville, AL, USA. 
\scriptsize$^{8}$NASA Postdoctoral Program Fellow, USA. \\
\scriptsize$^{9}$Astrophysics Group, Cavendish Laboratory, JJ Thomson Avenue, Cambridge. 
\scriptsize$^{10}$Instituto de Fisica de Cantabria (CSIC-UC), E39005-Santander, Spain. \\
\scriptsize$^{11}$Instituto de Astrof\' isica de Andaluc\' ia (IAA-CSIC), Granada, Spain. 
\scriptsize$^{12}$Dark Cosmology Centre, Niels Bohr Institute, University of Copenhagen, Copenhagen, Denmark.\\
\scriptsize$^{13}$CEOU, Dept. of Physics \& Astronomy, Seoul National University, Seoul, KOREA. \\
\scriptsize$^{14}$Astronomical Institute, University of Amsterdam, XH Amsterdam, The Netherlands.
\scriptsize$^{15}$Indian Institute of Astrophysics, Bangalore, India. \\
\scriptsize$^{16}$Observatori Astron\' omic de la Universitat de Val\` encia, Valencia, Spain. 
\scriptsize$^{17}$University of Sheffield, Department of Physics \& Astronomy, Sheffield, UK. \\
\scriptsize$^{18}$Deutsches SOFIA Institut, Universit\" ät Stuttgart, Stuttgart, Germany. 
\scriptsize$^{19}$SOFIA Science Center, NASA Ames Research Center, Moffett Field CA, USA. \\
\scriptsize$^{20}$AIM, CEA/DSM - CNRS - Universit\'e Paris Diderot, Irfu/SAP, Centre de Saclay, France. \\
\scriptsize$^{21}$Special Astrophysical Observatory of Russian Academy of Science (SAO-RAS), Nizhnij Arkhyz, Karachai-Cherkessia, Russia. \\
\scriptsize$^{22}$Space Telescope Science Institute, Baltimore, MD, USA. 
\scriptsize$^{23}$Physics Dept., University of Notre Dame, Notre Dame, IN, USA. \\
\scriptsize$^{24}$Faculty of Mathematics and Physics, University of Ljubljana, Slovenia. 
\scriptsize$^{25}$Centre of Excellence SPACE-SI, A\v sker\v ceva cesta 12, SI-1000 Ljubljana Slovenia. \\
\scriptsize$^{26}$Department of Physics and Astronomy, University of North Carolina, NC, USA. 
\scriptsize$^{27}$Academia Sinica Institute of Astronomy and Astrophysics, Taipei, Taiwan. \\
\scriptsize$^{28}$University of Bristol, H.H.Wills Physics Laboratory, Bristol, UK. 
\scriptsize$^{29}$Ulugh Beg Astronomical Institute, Tashkent, Uzbekistan. \\
\scriptsize$^{30}$Korea Astronomy and Space Science Institute, Daejeon, Korea. 
\scriptsize$^{31}$University of California, Santa Cruz, CA, USA. 
\scriptsize$^{32}$Hubble Fellow. \\
\scriptsize$^{33}$Institute of Solar-Terrestrial Physics, Irkutsk, Russia. 
\scriptsize$^{34}$Space Science Office, VP62, NASA/Marshall Space Flight Center, Huntsville, AL, USA. \\
\scriptsize$^{35}$Max-Planck-Institut f\"ur Astrophysik, Karl-Schwarzschild- Strasse 1, Garching, Germany. 
\scriptsize$^{36}$Scuola Normale Superiore, Piazza Cavalieri 7, Pisa, Italy. \\
\scriptsize$^{37}$INAF Oss. Astron. Padova, vicolo dellOsservatorio 5, Padova, Italy. 
\scriptsize$^{38}$Computational Cosmology Center, Lawrence Berkeley National Laboratory, Berkeley, CA, USA. \\
\scriptsize$^{39}$Osservatorio Astronomico Di Trieste, Via G.B.Tiepolo, Trieste, Italy. 
\scriptsize$^{40}$School of Earth and Space Exploration, Arizona State University, Tempe, AZ, USA. \\
\scriptsize$^{41}$SRI Crimean Astrophysical Observatory, Nauchny, Crimea, Ukraine. 
\scriptsize$^{42}$Carnegie Observatories, 813 Santa Barbara Street, Pasadena, California, USA. \\
\scriptsize$^{43}$ASTRON, Radio Observatory, Dwingeloo, Netherlands. 
\scriptsize$^{44}$Astronomy Centre, James Cook University, Townsville, Australia.  \\
\scriptsize$^{45}$Las Cumbres Observatory Global Telescope Network, Goleta, CA, USA. \\
\scriptsize$^{46}$Astronomy Unit, School of Mathematical Sciences, Queen Mary, University of London, London, UK. \\
\scriptsize$^{47}$Institute of Astronomy, National Central University, Chung-Li, Taiwan. \\
\scriptsize$^{48}$Isaac Newton Group of Telescopes, Apartado de correos 321, E-38700 Santa Cruz de la Palma, Canary Islands, Spain. \\
\scriptsize$^{49}$Instituto de Astrof\' isica de Canarias, 38200 La Laguna, Tenerife, Spain. 
\scriptsize$^{50}$Sternberg Astronomical Institute, Moscow State University, Moscow, Russia. \\
\scriptsize$^{51}$Astrophysics Research Centre, School of Mathematics and Physics, Queen's University Belfast, Belfast, UK.
}
\begin{document}

\maketitle

\begin{abstract}

We present ground-based and \textit{HST} optical observations of the optical transients (OTs) of long-duration Gamma Ray Bursts (GRBs) 060729 and 090618, both at a redshift of $z=0.54$.  For GRB 060729, bumps are seen in the optical light curves (LCs), and the late-time broadband spectral energy distributions (SEDs) of the OT resemble those of local type Ic supernovae (SNe).  For GRB 090618, the dense sampling of our optical observations has allowed us to detect well-defined bumps in the optical LCs, as well as a change in colour, that are indicative of light coming from a core-collapse SN.  The accompanying SNe for both events are individually compared with SN1998bw, a known GRB-supernova, and SN1994I, a typical type Ic supernova without a known GRB counterpart, and in both cases the brightness and temporal evolution more closely resemble SN1998bw.  We also exploit our extensive optical and radio data for GRB 090618, as well as the publicly-available \textit{SWIFT}-XRT data, and discuss the properties of the afterglow at early times.  In the context of a simple jet-like model, the afterglow of GRB 090618 is best explained by the presence of a jet-break at $t-t_{o}>0.5$ days.  We then compare the rest-frame, peak $V$-band absolute magnitudes of all of the GRB and X-Ray Flash (XRF)-associated SNe with a large sample of local type Ibc SNe, concluding that, when host extinction is considered, the peak magnitudes of the GRB/XRF-SNe cannot be distinguished from the peak magnitudes of non-GRB/XRF SNe.

\end{abstract}

\begin{keywords}
 gamma-ray burst: (individual: GRB 060729)---gamma-ray burst: (individual: GRB 090618)---supernovae: general.  
\end{keywords}

\section{Introduction}

Compelling evidence connecting the generation of bursts of gamma-rays and flashes of X-rays with the gravitational core-collapse of stripped-envelope supernovae (SNe) continues to grow, e.g., XRF 100316D and SN2010bh (Chornock et al. 2010; Starling et al. 2010).  The connection between GRB 980425 and unusual type Ic SN1998bw (Galama et al. 1998) provided an intriguing clue to the so-called ``GRB-SN'' connection (e.g., Woosley \& Bloom 2006) that was later confirmed with the ``smoking-gun'' detection of GRB 030329 and SN2003dh (Stanek et al. 2003; Hjorth et al. 2003; Matheson et al. 2003).  

Supporting the spectroscopic detections of long-duration GRBs (i.e., burst durations greater than $\geq 2$ s) and XRF-associated SNe are numerous photometric detections, e.g., XRF 020903 (Bersier et al. 2006); GRB 041006 (Stanek et al. 2005).  Late-time deviations from the power-law decline expected for a GRB afterglow, that are accompanied by a change in colour, are interpreted as evidence of supernovae (e.g., Zeh et al. 2004).  This spectroscopic and photometric evidence appears to strongly favour massive-star models, such as the ``Collapsar'' model (Woosley 1993) and the millisecond-magnetar model (e.g., Usov 1992; Thompson 1994; Wheeler et al. 2000; Zhang \& M\' esz\' aros 2001), for GRB production.

Indirect evidence supporting the ``GRB-SN'' connection was the realization that GRBs and type Ibc SNe occur in the brightest regions of their host galaxies (Fruchter et al. 2006; Kelly et al. 2008).  These observations link GRBs and XRFs to sites of star-formation, and consequently to massive stars.  Similar observations of the positions of Wolf-Rayet (WR) stars, a possible progenitor of GRBs and XRFs, also show that their distribution in their host galaxies favour the brightest regions (e.g., Leloudas et al. 2010).  Additionally, observations of GRB host galaxy morphology reveal a high fraction of merging and interacting systems (e.g., Conselice et al. 2005; Wainwright et al. 2007), which was shown to be an efficient trigger of star formation in galaxies (e.g., Joseph et al. 1984; Kennicutt \& Keel 1984).

The connection between long-duration GRBs and bright supernovae is not without complexity, however.  While questions surround the nature of GRB 980425 in its relation to ``classical'' GRBs, two nearby events have challenged the idea that \textit{all} long-duration GRBs are accompanied by an optically-bright SN. GRBs 060505 and 060614 (Gal-Yam et al. 2006; Fynbo et al. 2006; Della Valle et al. 2006; Ofek et al. 2007) had no accompanying supernova down to limits more than 100 times fainter than SN1998bw (though the classification of GRB 060614 as a long-duration GRB is questionable (e.g., Gehrels et al. 2006; Norris et al. 2010).  Thus the homogeneity of the ``GRB-SN'' connection, where the na\"ive expectation that \textit{all} long-duration GRB events produce enough nickel to power an optically-bright SN, may be an over-simplification of a much more complex reality.

So while the ``GRB-SN'' connection has been established, many questions still remain such as: ``What kind of progenitors produce these events?  Are the progenitors all the same?  Why do some massive stars become GRB/XRFs while most do not?''  General understanding of the types of progenitors that can give rise to a GRB, as well as the environments they occur in, have provided indirect clues.  The occurrence of ``optically dark bursts'' (i.e., GRBs without an optical afterglow) and the (sometimes) large amounts of dust in the vicinity of GRB progenitors that can obscure light from a supernova is a first attempt to explain the lack of SNe for some nearby GRBs, e.g., GRB 090417B (Holland et al. 2010).  Additionally, nearby extremely faint (more than 100 times fainter than SN1998bw) core-collapse SNe have been detected (Zampieri et al. 2003; Valenti et al. 2009\footnote{Though see Foley et al. (2009) and Foley et al. (2010) who present various lines of evidence that SN 2008ha was the result of a thermonuclear explosion of a carbon-oxygen white dwarf.}) with low expansion velocities and extremely low nickel production.  These under-luminous events provide another plausible explanation (e.g., Tominga et al. 2007; Moriya et al. 2010) for the lack of accompanying SNe with long-duration GRBs as well as illustrating the diverse optical properties of core-collapse SNe.

The evidence gathered so far shows that at least \textit{some} long-duration GRBs and XRFs are accompanied by an optically-bright type Ic SN.  Individual cases of spectroscopically-identified connections of type Ibc supernovae to GRBs and XRFs are the ideal way to further our understanding, however this is only technologically viable for low-redshift bursts.  Circumstantial evidence provided photometrically has also furthered our understanding by highlighting the enigmatic differences in GRB/XRF-associated SNe.  

Here we present results of data obtained on \textit{HST} and ground-based facilities of GRB 060729, and ground-based data for GRB 090618, both of which are at a common redshift of $z = 0.54$.  In section 2 we present our photometry and results for GRB 060729, effectively exploiting our \textit{HST} data to obtain image-subtracted magnitudes that provide evidence for an associated supernova.  In section 3 we present our photometry and results for GRB 090618, which resulted in densely-sampled $R_{c}$ and $i$ band light curves (LCs) that clearly show ``bumps'' that are accompanied by a change in colour that we attribute to flux coming from a stripped-envelope, core-collapse supernova.  In section 4 we compare our results of the rest-frame, absolute magnitudes of the SNe with those of: (1) all GRB/XRF-associated SNe and, (2) a sample of local type Ibc SNe, concluding that the progenitors that give rise to GRB/XRF SNe may be similar to those that produce local type Ibc SNe but with some differences.

Throughout the paper, observer-frame times are used unless specified otherwise in the text.  The respective decay and energy spectral indices $\alpha$ and $\beta$ are defined by $f_{\nu} \propto (t - t_{0})^{-\alpha}\nu^{-\beta}$, where $t_{0}$ is the time of burst and $\nu$ is the frequency.  Foreground reddening has been corrected for using the dust maps of Schlegel et al. (1998), from which we find $E(B-V) =0.055$ mag for GRB 060729 and $E(B - V ) = 0.085$ mag for GRB 090618.  We adopt a flat $\Lambda$CDM cosmology with a Hubble parameter of $H_{0} = 71$ km/s/Mpc, a matter density of $\Omega_{M} = 0.27$, and a cosmological constant of $\Omega_{\Lambda} = 1 - \Omega_{M} = 0.73$.  For this cosmology a redshift of $z = 0.54$ corresponds to a luminosity distance of $d_{L} = 3098.7$ Mpc and a distance modulus of $42.45$ mag.  

\section{GRB 060729}

The \textit{Swift} Burst Alert Telescope (BAT) detected the long-duration ($T_{90} \approx 115$ s; where $T_{90}$  ) GRB 060729 on July 29, 2006 (Grupe et al. 2006) with a remarkably bright and long-lasting X-ray afterglow (Grupe et al. 2007; Grupe et al. 2010).  The redshift was measured by Th\"one et al. (2006) to be $z = 0.54$, and a later spectroscopic analysis by Fynbo et al. (2009) measured the redshift to be $z=0.5428$.

\subsection{Observations \& Photometry}
\subsubsection{\textit{Ground-based Data}}

\begin{table*}
\caption{Ground-based Photometry of GRB 060729 \label{table:060729_ground}}
\begin{tabular}{cccccc}
\hline
$T-T_{o}$ (days) & Telescope$^{a}$ & Filter & Mag & $\sigma$ (mag) & Calibrated to \\
\hline
0.452 & Prompt & $B$ & 18.22 & 0.09 & Vega \\
0.452 & Prompt & $z$ & 17.36 & 0.08 & AB \\
0.481 & Prompt & $z$ & 17.29 & 0.07 & AB\\
0.515 & Prompt & $z$ & 17.22 & 0.06 & AB\\
0.551 & Prompt & $z$ & 17.49 & 0.07 & AB\\
0.571 & Prompt & $B$ & 18.40 & 0.04 & Vega \\
0.594 & Prompt & $z$ & 17.49 & 0.07 & AB\\
1.574 & Prompt & $z$ & 18.43 & 0.17 & AB\\
4.601 & Gemini-S & $R_{c}$ & 20.24 & 0.04 & Vega \\
4.613 & Gemini-S & $g$ & 20.91 & 0.04 & AB\\
4.626 & Gemini-S & $I_{c}$ & 19.92 & 0.03 & Vega \\
4.638 & Gemini-S & $z$ & 20.51 & 0.04 & AB\\
17.598 & CTIO & $R_{c}$ & 21.90 & 0.06 & Vega \\
46.583 & Gemini-S & $R_{c}$ & 23.29 & 0.06 & Vega \\
49.554 & Gemini-S & $R_{c}$ & 23.32 & 0.05 & Vega \\
\hline
\end{tabular}

\medskip
$^{a}$Telescope key: CTIO: $4$m Cerro Tololo Blanco Telescope; Prompt: $0.41$m Prompt Telescope; Gemini-S: $8.1$m Gemini-South Telescope.
All magnitudes have been corrected for foreground extinction.
\end{table*}

We obtained data with four ground-based telescopes: Panchromatic Robotic Optical Monitoring and Polarimetry Telescopes (PROMPT), Cerro Tololo Inter-American Observatory (CTIO), Faulkes Telescope South (FTS) \& Gemini-South.  Two epochs of data were taken with one of the $0.41$m Panchromatic Robotic Optical Monitoring and Polarimetry Telescopes (PROMPT) $\sim 0.5$ \& $1.5$ days after the burst.  The first epoch yielded images in $B$ and $z$, while the second epoch was only in $z$.  One epoch of data was obtained on the Cerro Tololo (CTIO) Blanco $4$m telescope $17.6$ days after the burst, yielding an $R_{c}$ image.  Three epochs of data were obtained on the Gemini-South (Gemini-S) $8.1$m telescope, the first one was $4.6$ days post-burst in $griz$, and two additional epochs in $r$ at $46.5$ and $49.5$ days post-burst.  Finally, one epoch of images taken was obtained on $5^{th}$ April, 2007 ($\sim 250$ days after the burst) on the $2$m Faulkes Telescope South (FTS).  Observations were made of the GRB field (well after the GRB and SN had faded below the instrument detection limit) in $BVR_{c}i$, as well as images taken of Landolt photometric standard regions (Landolt 1992) in the same filters.  The observations taken by FTS that are used in our calibration were made under photometric conditions.

Aperture photometry was performed on all images using standard routines in IRAF\footnote{IRAF is distributed by the National Optical Astronomy Observatory, which is operated by the Association of Universities for Research in Astronomy, Inc., under cooperative agreement with the National Science Foundation.}.  A small aperture was used, and an aperture correction was computed and applied.  The aperture-corrected, instrumental magnitudes were then calibrated via standard star photometry into magnitudes in $BVR_{c}I_{c}$.  Using the images of Landolt standards, IRAF routines were used to solve transformation equations of the form:

\begin{equation}
 m_{inst}\ =\ z_{p}\ +\ M\ +\ a_{1}X\ +\ a_{2}(B-V) 
\end{equation}

{\noindent}where $m_{inst}$ is the aperture-corrected, instrumental magnitude, $M$ is the standard magnitude, $z_{p}$ the zero-point, $a_{1}$ the extinction, $X$ the airmass, $a_{2}$ the colour-term and $(B-V)$ (and variations thereof) the colour.  

The validity of the transformation equations was checked by using the solutions on the Landolt standard stars in the CCD images, which revealed computed magnitudes that were compatible with those in the Landolt Catalogue (1992) within the magnitude errorbars (typical $\sigma \sim 0.04$ mag).  

The solutions from the standard star photometry were then applied to a sequence of secondary standards in the field of GRB 060729.  The Gemini-S observations are calibrated against these stars using a zero-point and colour term in filters $gR_{c}I_{c}z$.   The $g$ and $z$ magnitudes of secondary standards in the GRB field were calculated using transformation equations from Jordi et al. (2006), and $r$ and $i$ were calibrated to $R_{c}$ and $I_{c}$.  The single CTIO epoch was calibrated directly against the stars in the GRB field using a zero-point (no colour term).  Images taken by PROMPT are calibrated against standard stars observed by PROMPT in $B$ and $z$, using a zero-point and colour term.  All of the ground-based photometry, which has been corrected for foreground extinction, is presented in Table \ref{table:060729_ground}.

\subsubsection{\textit{Hubble Space Telescope Data}}

\begin{table}
\caption{\textit{HST} Photometry of GRB 060729 \label{table:060729_HST}}
\begin{tabular}{ccccc}
\hline
$T-T_{o}$ (days) & Filter$^{a}$ & Mag$^{b}$ & $\sigma$ (mag) & Calibrated to \\
\hline
8.983 & $F330W$ & 22.00 & 0.03 & AB \\
17.051 & $F330W$ & 23.40 & 0.05 & AB \\
26.125 & $F330W$ & 24.08 & 0.10 & AB \\
9.042 & $F625W$ & 21.09 & 0.01 & $R_{c}$ (Vega) \\
16.909 & $F625W$ & 21.94 & 0.02 & $R_{c}$ (Vega) \\
26.250 & $F625W$ & 22.57 & 0.03 & $R_{c}$ (Vega) \\
48.689 & $F625W$ & 24.07 & 0.07 & $R_{c}$ (Vega) \\
139.888 & $F625W$ & 26.20 & 0.62 & $R_{c}$ (Vega) \\
9.059 & $F850LP$ & 20.90 & 0.01 & $I_{c}$ (Vega) \\
16.926 & $F850LP$ & 21.74 & 0.02 & $I_{c}$ (Vega) \\
26.266 & $F850LP$ & 22.24 & 0.03 & $I_{c}$ (Vega) \\
48.757 & $F850LP$ & 23.28 & 0.05 & $I_{c}$ (Vega) \\
139.905 & $F850LP$ & 25.65 & 0.57 & $I_{c}$ (Vega) \\
9.139 & $F160W$ & 21.14 & 0.02 & AB \\
16.500 & $F160W$ & 22.16 & 0.04 & AB \\  
27.198 & $F160W$ & 22.85 & 0.05 & AB \\  
9.124 & $F222M$ & 20.38 & 0.30 & AB \\
16.460 & $F222M$ & $>$ 22.1 & - & AB \\
27.127 & $F222M$ & $>$ 21.7 & - & AB \\
\hline
\end{tabular}

\medskip
$^{a}$Filter key: $F330W$, $F625W$, $F850LP$: ACS; $F160W$, $F222M$: NICMOS. \\
$^{b}$Apparent magnitudes are of the OT from the subtracted images \textit{except for} the ACS $F330W$ and NICMOS $F222M$ filters (for which no image subtraction occurred).\\
All magnitudes have been corrected for foreground extinction.
\end{table}

We obtained a total of six epochs of data with \textit{HST}, using the Advanced Camera for Surveys (ACS) and the High-Resolution Channel (HRC) in $F330W$; and the Wide Field Channel (WFC) in $F625W$ and $F850LP$; the Near Infrared Camera and Multi-Object Spectrograph (NICMOS) in $F160W$ and $F222M$; and the Wide-Field Planetary Camera 2 (WFPC2) in $F300W$, $F622W$ and $F850LP$ for the last epoch.  The images were reduced and drizzled in a standard manner using \textit{Multidrizzle} (Koekemoer et al. 2002).  The use of WPFC2 images in the last epoch was a consequence of the ACS being broken at this time.  While we were wary that images from two different cameras and two different filters might disrupt our intention of using the last epoch as a template image for use in image subtraction, we were able to work around the problem of the differing cameras and filters by carefully choosing the stars used in the image-subtraction procedure (see section 2.1.3).  The ACS images that were used for image subtraction were drizzled to the WFPC2 scale ($0.10''$), while the remaining images were drizzled to the native scale of the individual cameras ($0.2''$ and $0.10''$ for NICMOS and WFPC2 respectively).

Image subtraction was not carried out on the $F330W$ images because the host is not visible in the final WFPC2 ($F300W$) image, thus we used the procedure described in Appendix B of Sirianni et al. (2005) for the ACS data: (a) aperture photometry using aperture of $0.15''$, (b) aperture correction for apertures from $0.15''$ to $0.50''$, (c) aperture correction for $F330W$ for aperture $0.50''$ to infinity (Table 5 of Sirianni et al. 2005), (d) CCD charge transfer efficiency (CTE), (e) used AB zeropoint.  

\subsubsection{\textit{Image Subtraction}}

Using ISIS (Alard 2000) we performed image subtraction on the ACS $F625W$ and $F850LP$ images, using the WFPC2 images in $F622W$ and $F850LP$ as the respective templates.  The ACS images were drizzled to the WFPC2 scale ($0.10''$) prior to subtraction.  Image subtraction was also performed on the NICMOS $F160W$ images, using the last epoch as the template.  Subtracting the WFPC2 epoch from the first five ACS epochs gave a clear detection of the optical transient (OT).  As the reference image was taken with a different camera and different filters (with the corresponding differences in the filters' transmission curves and different CCD responses) we checked for the presence of a colour-term in the residuals in the subtracted images.  For very red objects ($F622W-F850LP > 1.5$) a small colour dependent effect was seen, however when we restricted the colour range to include only stars with $0 \leq F622W-F850LP \leq 1.0$ we found no statistically-significant colour-term in the residuals.  The colour of the OT was never beyond this colour range, thus by using the stars in this colour range to match the ACS frames to the WFPC2 template in our image-subtraction procedure we are confident of the validity of our results. 

Aperture photometry was then performed on the subtracted images using an aperture of $0.15''$, and correcting for steps (a) - (c) as described previously but for the WFPC2 images (Holtzman et al. 1995).  The instrumental magnitudes were then calibrated to $R_{c}$ and $I_{c}$ using a zero-point and colour-term.  Aperture photometry was performed on the NICMOS $F160W$ images using the procedures prescribed in the NICMOS data handbook (Thatte et al. 2009).  We used an aperture of $0.40''$ and corrected for: (a) aperture correction for apertures from $0.40''$ to $1.00''$, (b) correct to infinite aperture by multiplying the flux (counts) by $1.075$, (c) used AB zeropoint.  All of the \textit{HST} magnitudes, which have been corrected for foreground extinction, are listed in Table \ref{table:060729_HST}.

\subsubsection{\textit{Host Photometry}}

\begin{table}
\caption{\textit{HST} Photometry of the Host Galaxy of GRB 060729 \label{table:060729_HST_host}}
\begin{tabular}{ccccc}
\hline
$T-T_{o}$ (days) & Filter$^{a}$ &  Mag  & $\sigma$ (mag) & Calibrated to \\
\hline
425.298 & $F300W$ & $>$ 24.80 & - & AB \\
428.984 & $F622W$ & 24.42  & 0.20 & $R_{c}$ (Vega) \\
429.181 & $F850LP$ & 24.26  & 0.20 & $I_{c}$ (Vega)  \\
425.363 & $F160W$ & 22.82 & 0.06 & AB \\  
424.806 & $F222M$ & $>$ 22.10 & - & AB \\ 
\hline
\end{tabular}

\medskip
$^{a}$Filter key: $F622W$ \& $F850LP$: WFPC2; $F160W$: NICMOS.\\
All magnitudes have been corrected for foreground extinction.
\end{table}

The host was clearly visible and extended in all of the final epoch images taken by WFPC2, apart from the $F300W$ image.  Aperture photometry was performed on the images: WFPC2 ($F622W$ and $F850LP$) and NICMOS ($F160W$), using an aperture of $0.7''$, $0.4''$ and $0.6''$ respectively.  As the host is clearly extended, CTE effects have been neglected when determining the host magnitudes as the effects due to CTE on WFPC2 images is only well understood for point-sources.  The corrected instrumental magnitudes in $F622W$ and $F850LP$ were then calibrated to $R_{c}$ and $I_{c}$ using a zero-point and colour-term, while the $F160W$ and $F222M$ magnitudes are given as AB magnitudes.  All of the host magnitudes,  which have been corrected for foreground extinction, are listed in Table \ref{table:060729_HST_host}.

\subsection{Results \& Discussion}

\subsubsection{The Afterglow}

\begin{figure}
\centering
\includegraphics[height=3.4in,width=2.5in, angle=270]{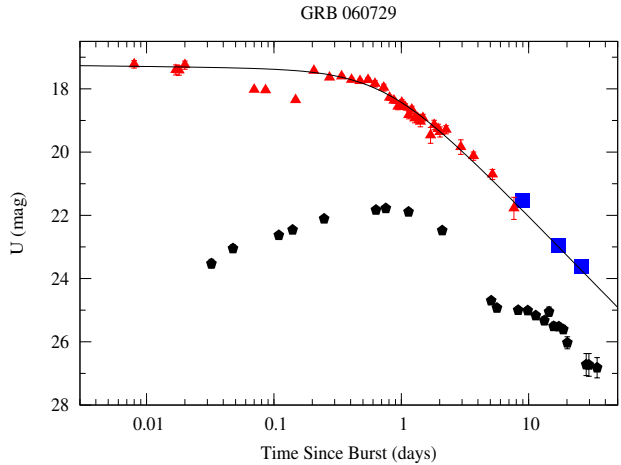}
\caption{GRB 060729: Light-curve of \emph{UVOT (U)} (triangles) and \emph{HST (F330W)} (squares) magnitudes transformed into $U$.  The afterglow model is shown as a broken power-law (solid line), where $\alpha_{1} = 0.01 \pm 0.03$; $\alpha_{2} = 1.65 \pm 0.05$ and $T_{break} = 0.75 \pm 0.08$ days.  Plotted for comparison (pentagons) is the \textit{UVOT} ($UVW1$) LC of SN 2006aj/GRB 060218 (Brown et al. 2009) as it would appear at $z=0.54$.   All magnitudes have been corrected for foreground extinction and the LC of 2006aj has also been corrected for host extinction.  At these wavelengths the contribution of flux from an accompanying SN is expected to be negligible in comparison to the afterglow.  This assumption is confirmed by the $HST$ observations.}
\label{fig:060729_U}
\end{figure}

Grupe et al. (2007) found that the optical and X-ray light curves of GRB 060729 displayed similar temporal behaviour (i.e., the decay constants and time of break in the X-ray, UV and optical filters were consistent up to late times; see Figure 8 in Grupe et al. 2007).  Thus, assuming at late times that the decay constant will be approximately the same in all optical passbands, we used our \textit{HST}(ACS) $F330W$ magnitudes to characterise the temporal behaviour of the GRB afterglow at longer wavelengths (i.e., $R_{c}$ and $I_{c}$).  The reason for using the $F330W$ magnitudes is that any SN contribution at these wavelengths is expected to be negligible with respect to the afterglow (e.g., see Figure \ref{fig:060729_U}).

Using the Ultraviolet and Optical Telescope (UVOT) aboard \textit{Swift}, Grupe et al. (2007) found that the optical afterglow was visible in the UVOT-$U$ filter up to $7.66$ days after the burst.  We clearly detect the GRB afterglow up to our third \textit{HST} epoch ($26.12$ days after the burst) in the \textit{HST}(ACS) $F330W$ filter.  Thus, by combining our \textit{HST} data with that obtained by \textit{Swift}, we were able to extend the time-domain of the analysis and more accurately determine the temporal behaviour of the afterglow.

To enable us to compare the $F330W$ magnitudes obtained with \textit{HST}(ACS) to those determined by \textit{Swift}, we used IRAF/Synphot/Calcphot to transform our $F330W$ magnitudes into $U$ and corrected for foreground extinction.  Next, using transformation equations determined by \textit{Swift} (Poole et al. 2008), we transformed the UVOT-$U$ magnitudes into $U$.  All of the $U$-band data (\textit{HST} and \textit{Swift}) were thus calibrated to $U$.

A broken power-law was fitted to the data (Figure \ref{fig:060729_U}), using the form:

\begin{equation}
 m_{\nu}(t) = -2.5 \times \log \left(\left(\left(\frac{t}{T_{break}}\right)^{\alpha_{1}} + \left(\frac{t}{T_{break}}\right)^{\alpha_{2}}\right)^{-1}\right) + B
\end{equation}

{\noindent}where $t$ is the time since the burst and $T_{break}$ is the time when the power-law changes from temporal index $\alpha_{1}$ to $\alpha_{2}$.

The values of the fit ($\chi^{2}  /  dof =  1.31$) are: $\alpha_{1} = 0.01 \pm 0.03$; $\alpha_{2} = 1.65 \pm 0.05$ and $T_{break} =  0.75 \pm 0.08$ days.  When we restrict the upper limit to the time range to $8$ days (i.e., up to the last UVOT-$U$ detection) we find ($\chi^{2}  /  dof =  1.33$): $\alpha_{2} =  1.47 \pm 0.11$ and $T_{break} =  0.58 \pm 0.10$ days, which are consistent with the decay constant and time of break found by Grupe et al. ($2007$) for the UVOT-$U$ data (see Table $5$ of Grupe et al. $2007$).  We note that we originally fit the data with the complete Beuermann Function (Beuermann et al. 1999) and let $n$ vary (where $n$ is a measure of the ``smoothness'' of the transition from $\alpha_{1}$ to $\alpha_{2}$).  The results of our fit gave $n \approx 1$ (i.e., a smooth transition), so we fixed $n=1$ when performing the final fit.

In Figure \ref{fig:060729_U}, variability is seen in the early-time data.  To check that the early-time data does not affect our empirical fit at late-times, we fit a single power-law to the data at $t-t_{o} > 2.0$ days.  It was seen that by limiting the fit to data after this time, we found $\alpha = 1.64 \pm 0.07$, fully consistent with the value of $\alpha_{2}$ found previously.

The value of $\alpha_{2} = 1.65 \pm 0.05$ is consistent with that found by Grupe et al. (2010) for the X-ray light curve ($\alpha = 1.61^{+0.10}_{-0.06}$) from $1.2 \times 10^{6}$ s ($13.8$ days) to $38 \times 10^{6}$ s ($439.8$ days), though we note that the optical and X-ray decay constants do not necessarily \textit{have} to be the same, as seen for many bursts (e.g., Melandri et al. 2008).

Plotted for comparison in Figure \ref{fig:060729_U} is the \textit{UVOT (UVW1)} LC of XRF 060218/SN 2006aj. At a redshift of $z=0.54$, the light emitted in the rest-frame at \textit{UVOT (UVW1)} wavelengths will be redshifted into observer-frame $U$.  We can see that the broken power-law fits the $U$ data well and we conclude that the contribution of flux at this wavelength from an accompanying SN is negligible in comparison to the flux from the afterglow.

\subsubsection{The Supernova}

For the complete GRB event, we assume that flux is coming from up to three sources: the afterglow, the host and a (possible) SN.  We have already photometrically removed the flux due to the host in the \textit{HST} images via image subtraction, however we were not able to perform image subtraction on the Gemini-S and CTIO images.  Instead we mathematically subtracted the contribution of host flux from the three Gemini-S and CTIO images, after all of the magnitudes were converted into fluxes using the zero-points from Fukugita et al. (1995).  Thus, after removing the flux due to the host, and then subtracting the flux due to the GRB afterglow (which was characterised by Equation 2), any remaining flux could be due to an accompanying SN.

\begin{figure}
\centering
\includegraphics[height=3.4in,width=2.5in, angle=270]{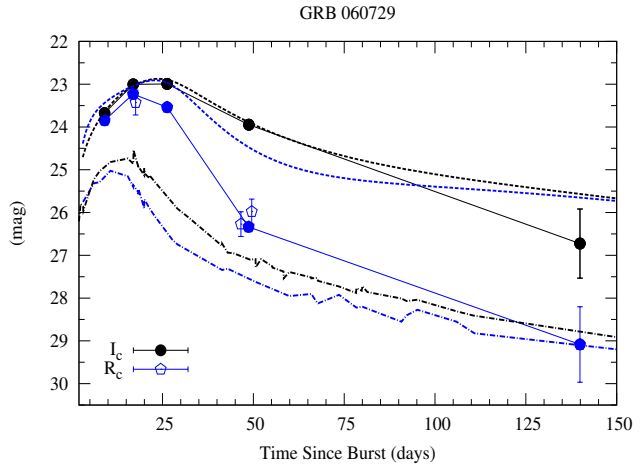}
\caption{GRB 060729:  afterglow-subtracted, supernova $R_{c}$ (blue) and $I_{c}$ (black) light curves.  The solid points are derived from the image-subtracted magnitudes from the \emph{HST} data while the open points are derived from the magnitudes obtained by mathematically subtracting the host flux from the Gemini-S and CTIO epochs (see text).  Plotted for comparison are the LCs of SN1998bw (dashed) and SN1994I (dot-dashed) in the same colours, as they would appear at $z=0.54$.  The brightness and time-evolution of the SN resembles that of SN1998bw rather than SN1994I.  All magnitudes have been corrected for foreground and host extinction.   }
    \label{fig:060729_SNe}
\end{figure}

Our host- and afterglow-subtracted light curves are shown in Figure \ref{fig:060729_SNe}, where the quoted errors have been calculated in quadrature.  Plotted for comparison are the distance and foreground-corrected light curves of SN1998bw and SN1994I.  The SN1998bw LCs are \textit{model} light curves (i.e., interpolated/extrapolated through frequency) however we have used the rest-frame $B$ and $V$ LCs of SN1994I to represent the observer-frame $R_{c}$ and $I_{c}$ LCs respectively.  It is seen, even despite the paucity of data-points, that the SN associated with GRB 060729 resembles more closely that of SN1998bw than SN1994I in regards to both time evolution and brightness.

We find peak apparent magnitudes of the associated SN of $R_{c} = 23.80 \pm 0.08$ and $I_{c} = 23.20 \pm 0.06$.  To make an accurate estimate of the peak absolute magnitude of the SN, we have determined the amount of host extinction (Section 2.2.4) to be $E(B-V) = 0.61$ mag.  In comparison, Grupe et al. (2007) determined from their optical to X-ray spectral energy distribution (SED) analysis of the afterglow a rest-frame extinction local to the GRB of $E(B-V) = 0.34$ mag.  However, using the same $UVOT$ data-set as Grupe et al. (2007) as well as additional $R_{c}$ data, Schady et al. (2010) found a rest-frame extinction of $E(B-V)\le 0.06$.  We note that our value of the host reddening is poorly constrained by our SED template fitting (though it is worth noting that the best two fits to the host SED both imply large amounts of reddening).  However, it has been seen that the host extinction for many bursts is not always the same as the extinction local to the event (e.g., GRB 000210 (Gorosabel et al. 2003), where the local reddening was very high, but the host extinction was negligible.  Further examples are seen in Modjaz et al. 2008 and Levesque et al. 2010).  Thus it is more reasonable to use the extinction derived by Schady et al. (2010) as it is local to the event and is a more complete analysis than that performed by Grupe et al. (2007).  

As the majority of long-duration GRB host galaxies contain dust that is Small Magellanic Cloud (SMC)-like (e.g., Starling et al. 2007; Schady et al. 2007; Kann et al. 2010), we applied the SMC extinction law ($R_{V} = 2.93$; Pei 1992) to the value of $E(B-V)$ found by Schady et al. (2010) to estimate the rest-frame extinction.  We then approximated the rest-frame $V$-band magnitudes from the observer-frame $I_{c}$-band magnitudes, and find an absolute magnitude of the SN associated with GRB 060729 of $M_{V} = -19.43 \pm 0.06$.  This implies that the SN associated with GRB 060729 has approximately the same peak brightness as 1998bw in the $V$-band (see Table \ref{table:GRB_SNe}).

\subsubsection{The Spectral Energy Distribution}

\begin{figure}
\centering
\includegraphics[height=3.4in,width=2.5in, angle=270]{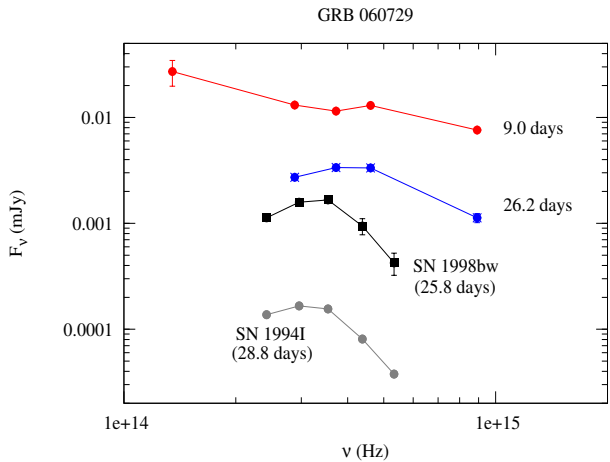}
\caption{GRB 060729: Observer-Frame SED.  The host-subtracted \textit{HST} data (filled circles) of the OT (afterglow and supernova) has been corrected for foreground and host extinction.  Plotted for comparison are the (see text) \textit{UBVRI} data of SN1998bw at $25.8$ days and the \textit{UBVRI} data for SN1994I at $28.8$ days, which have been corrected for foreground and host extinction.  At $9.0$ days the SED is fitted by a power-law with spectral index $\beta = 0.59 \pm 0.02$, however at $26.2$ days the SED resembles that of the local Ic SNe, although with less curvature. }
\label{fig:060729_SED}
\end{figure}

We determined the observer-frame spectral energy distribution of the OT (afterglow and SN) for two epochs: $9.0$ and $26.2$ days after the burst (Figure \ref{fig:060729_SED}).  We used the zero-points from Fukugita et al. (1995) to convert the magnitudes to fluxes, and corrected for foreground and host extinction.  Plotted for comparison are the observer-frame SEDs for SN1998bw ($25.8$ days) and SN1994I ($28.8$ days), which have been corrected for distance as well as foreground and host extinction.  At $9.0$ days the SED is fitted by a power-law with spectral index $\beta= 0.59 \pm 0.02$.  However, the most striking feature of Figure \ref{fig:060729_SED} is at $26.2$ days, where the SED resembles that of 1998bw and 1994I at a similar epoch, although with less curvature.

\subsubsection{The Host Galaxy}

\begin{figure}
\centering
\includegraphics[height=3.4in,width=3.4in, angle=0]{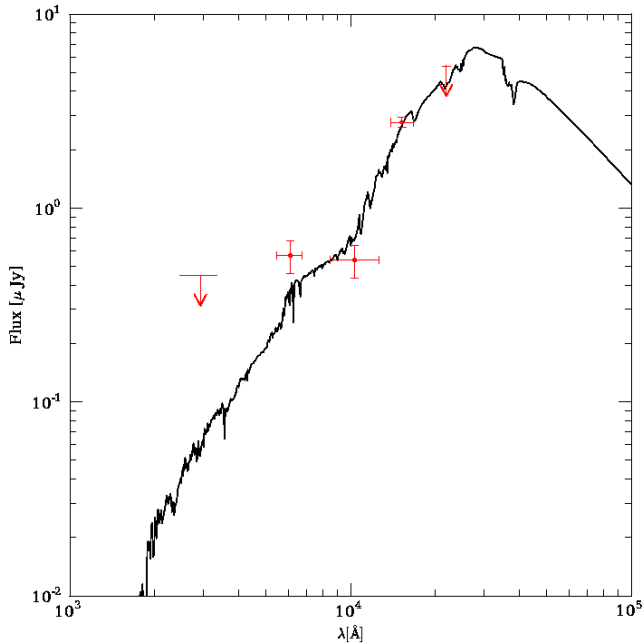}
 %\vspace{0.4in}
\caption{GRB 060729: SED of the host galaxy.  The best-fit model is for a galaxy with a young stellar population and low metallicity.}
    \label{fig:060729_host}
\end{figure}

The detection of the host galaxy with WFPC2 allowed us to construct its optical SED (Figure \ref{fig:060729_host}).  Magnitudes of the host, which are corrected for foreground reddening, were transformed to AB magnitudes, giving: $F622W = 24.54 \pm 0.20$, $F850LP = 24.60 \pm 0.20$, $F160W = 24.12 \pm 0.06$\footnote{A recent image of the host galaxy has been taken with the Wide Field Camera 3 aboard $HST$ in filter $F160W$ by A. Levan, who confirms that the host brightness in this filter has not changed between this epoch and our epoch taken at $\approx 425$ days (A. Levan, private communication).}, and upper limits: $F330 > 24.80$, $F222M > 22.10$ mag.

We used the procedure described in Svensson et al. (2010) to calculate the stellar masses, star-formation rate and the specific star-formation rate of the host galaxy.  Using a redshift of $z = 0.54$, we fitted our \textit{HST} magnitudes to synthetic galaxy evolution models (Bruzual \& Charlot 2003).  None of the synthetic models were very well constrained, however the best fit template ($\chi^{2}  /  dof =  6.58$) is for a galaxy with a young stellar population and low metallicity.  We note that the next best-fit model is for a M82-type galaxy with large amounts of reddening.  

The magnitudes of the best-fitting template are: $M_{U} = -16.72$, $M_{B} = -17.13$, $M_{V} = -17.49$, $M_{K} = -19.94$ and $E(B-V) \approx 0.61$ mag.  The magnitudes are not host-extinction corrected.  We also find a star-formation rate of $\log(SFR[\rm{M_{\sun}/yr}]) = -0.89^{+0.04}_{-0.07}$; a stellar mass of $\log(Mass[\rm{M_{\sun}}]) = 9.13^{+0.04}_{-0.08}$ and a specific star-formation rate of $\log(SSFR[\rm{Gyr^{-1}}]) = -1.03^{+0.18}_{-0.04}$.  The latter values of the star-formation rates and mass are estimated from an extinction-corrected SED.  The quoted errors are $1\sigma$ and are systematic only and are calculated from the distributions of SFR, mass, etc., given by all templates (i.e., the errors reflect how good the best-fit parameters are compared with all of the considered models).

We were able to determine the position of GRB 060729 in the host galaxy using the \textit{HST} images taken by WFPC2.  We find for the position of the OT in our subtracted images (error $=0.02''$): $06^{h}21^{m}31.77^{s}$, $-62^{d}22^{m}12.21^{s}$; which is offset from the apparent centre of the host by $ 0.33'' \pm 0.02''$ ($2.1 \pm 0.1$ kpc).  

Secondly, we used the pixel statistic created by Fruchter et al. (2006), which calculates the fraction of light ($F_{light}$) contained in regions of lower surface brightness than the region containing a GRB or SN (where a value of $F_{light} = 100.0$ corresponds to the event occuring in the brightest region of the host galaxy).  For GRB 060729, $F_{light} = 39.0$ in the $F622W$ filter, which is somewhat smaller than that seen in other GRBs by Fruchter et al. (2006).

Thus we conclude that the host galaxy of GRB 060729 has a young stellar population with a modest stellar mass ($\sim 10^{9}\ \rm{M_{\sun}}$) and star-formation rate ($SFR \sim 0.13\ \rm{M_{\sun}yr}^{-1}$), suggesting that the host is typical of other GRB host galaxies (e.g., Wainwright et al. 2005; Fruchter et al. 2006; Savaglio et al. 2009; Svensson et al. 2010; Christensen et al. 2004).

\section{GRB 090618}

The \textit{Swift} Burst Alert Telescope (BAT) detected the long-duration ($T_{90} = 113.2 \pm 6\ s$) GRB 090618 on June 18, 2009 at 08:28:29 UT (Schady et al. 2009) which was followed up by many teams with ground-based telescopes at optical wavelengths, and published in the GCN Circulars.\footnote{The data presented in this paper supersedes those in the GCN Circulars: FTN: Melandri et al. 2009 and Cano et al. 2009; LOAO: Im et al. 2009a;  OAUV-0.4m telescope: Fernandez-Soto et al. 2009; Shajn: Rumyantsev \& Pozanenko 2009; BOAO: Im et al. 2009b; SAO-RAS: Fatkhullin et al. 2009 (photometry \& spectra); HCT: Anupama, Gurugubelli \& Sahu 2009; Mondy: Klunko, Volnova \& Pozanenko 2009.  GRB 090618 was also detected at radio wavelengths and published in GCN Circulars: AMI: Pooley 2009a, 2009b; VLA: Chandra \& Frail 2009; WSRT: Kamble, van der Horst \& Wijiers 2009.}

The redshift was measured to be $z = 0.54$ by Cenko et al. (2009) and Fatkhullin et al. (2009).

\subsection{Observations \& Photometry}

\subsubsection{Optical Data}

\begin{table}
\caption{Ground-based Optical Photometry of GRB 090618$^{a}$ \label{table:090618_ground}} 
\begin{tabular}{ccccc}
\hline
$T-T_{o}$(days) & Filter &  Mag$^{b}$ & $\sigma$ (mag) & Telescope$^{c}$ \\
\hline
0.057	& $B$ & 	17.34	&	0.05	&	FTN	\\
0.059	& $B$ & 	17.38	&	0.05	&	FTN	\\
0.062	& $B$ & 	17.43	&	0.05	&	FTN	\\
0.067	& $B$ & 	17.50	&	0.06	&	FTN	\\
...	& ... & 	...	&	...	&	...	\\
\hline
\end{tabular}

\medskip
$^{a}$Only a section of the data is shown here.  The complete table is available online at: \\
$^{b}$The apparent magnitude of the GRB+SN+HOST.\\
$^{c}$Telescope key: AZT-11: $1.25$m AZT-11 Telescope of CrAO; BOAO: the $15.5$cm BOAO Telescope; BTA-6: $6$m SAO-RAS Telescope FTN: $2$m Faulkes Telescope North; Gemini (N): $8.1$m Gemini North; HCT: $2$m Himilayan Telescope; INT: $2.5$m Isaac Newton Telescope; LOAO: $1.0$m LOAO Telescope; LT: $2$m Liverpool Telescope; Maidanak: $1.5$m Maidanak Telescope; Mondy: $1.5$m AZT-33IK Telescope; OAUV: $0.4$m OAUV Telescope; Shajn: $2.6$m Shajn Telescope of CrAO; WHT: $4.2$m William Herschel Telescope.
\end{table}

We obtained optical observations with many ground-based telescopes: the $2$m Faulkes Telescope North; the $2$m Liverpool Telescope; the $8.1$m Gemini-North; the $4.2$m William-Herschel Telescope (WHT); the $2.5$m Isaac Newton Telescope (INT); the $2.6$m Shajn telescope of CrAO; the $1.25$m AZT-11 telescope of CrAO; the $1.5$m Maidanak AZT-22 telescope with the SNUCAM CCD camera (Im et al. 2010); the $1.5$m AZT-33IK telescope of Sayan observatory, Mondy; the $6$m BTA-6 optical telescope and the $1$m Zeiss-1000 telescope of the Special Astrophysical Observatory of the Russian Academy of Sciences (SAO-RAS); the $1$m LOAO telescope (Lee, Im \& Urata 2010) at Mt. Lemmon (Arizona, USA) which is operated by the Korea Astronomy Space Science Institute; $15.5$cm telescope at Bohyunsan Optical Astronomy Observatory (BOAO) in Korea which is operated by the Korea Astronomy Space Science Institute; the $2$m Himalayan Chandra Telescope (HCT) and the T40 telescope at the Observatorio de Aras de los Olmos, operated by the Observatori Astron\' omic de la Universitat de Val\` encia.

Aperture photometry was performed on all images using standard routines in IRAF.  A small aperture was used, and an aperture correction was computed and applied.  The aperture-corrected, instrumental magnitudes of stars in the GRB field were calibrated to the \textit{SDSS} catalogue (in filters $BVR_{c}i$, where non-SDSS magnitudes were calculated using transformation equations from Jordi et al. (2006) using Equation 1, initially using a zero-point and a colour term.  However, it was found that the contribution of a colour term to the calibration was negligible, hence only a zero-point was used in all cases.  All optical magnitudes have been corrected for foreground extinction and are listed in Table \ref{table:090618_ground}.

\subsubsection{X-ray Data}

The X-ray afterglow was detected and monitored by \textit{Swift}-XRT (Schady et al. 2009; Evans et al. 2009; Beardmore \& Schady 2009).  The XRT data have been processed with the \textsc{heasoft} package v.6.9 and the corresponding calibration files (standard filtering and screening criteria have been applied).  For $t \la 240$ s the data are strongly affected by pile-up: piled-up Window Timing (WT) mode data have been corrected by eliminating a strip of data from the original rectangular region of event extraction: the strip size has been estimated from the grade 0 distribution (according to the \textit {Swift} nomenclature: see Burrows et al. 2005; Romano et al. 2006). Piled-up Photon Counting (PC) data have been extracted from an annular region whose inner radius has been derived comparing the observed to the nominal point spread function (PSF, Moretti et al. 2005; Vaughan et al. 2006).  The background is estimated from a source-free portion of the sky and then subtracted. The 0.3-10 keV background-subtracted, PSF- and vignetting-corrected light curve has been re-binned so as to ensure a minimum number of 25 re-constructed counts.  The count-rate light curves are calibrated into flux and luminosity light curves using a time dependent count-to-flux conversion factor as detailed in Margutti et al. (2010). In this way the strong spectral evolution detected at $t \la 240$ s is properly accounted for.

\subsubsection{Radio Data}

\begin{table*}
\begin{minipage}{175mm}
\caption{Radio Observations$^{a}$ of GRB 090618  \label{table:090618_radio}}
\begin{tabular}{cccccccccccc}
\hline
$T-T_{o}$ (days) & $F_{2.3\ GHz}$  &  $\sigma$  & $F_{4.8\ GHz}$  &  $\sigma$  & $F_{8.46\ GHz}$  & $\sigma$ & $F_{15\ GHz}$  & $\sigma$ & $F_{22\ GHz}$  &  $\sigma$  & Telescope$^{b}$ \\
\hline

0.168	&	-	&	-	&	-	&	-	&	-	&	-	&	-	&	-	&	$<$5	&	-	&	RT-22	\\
0.824	&	-	&	-	&	-	&	-	&	-	&	-	&	0.620	&	0.115	&	-	&	-	&	AMI	\\
0.980	&	-	&	-	&	-	&	-	&	0.383	&	0.047	&	-	&	-	&	-	&	-	&	VLA$^{c}$	\\
1.680	&	-	&	-	&	0.388	&	0.030	&	-	&	-	&	-	&	-	&	-	&	-	&	WSRT	\\
3.203	&	$<$3	&	-	&	-	&	-	&	$<$3	&	-	&	-	&	-	&	-	&	-	&	RT-22	\\
4.670	&	-	&	-	&	0.093	&	0.045	&	-	&	-	&	-	&	-	&	-	&	-	&	WSRT	\\
4.700	&	0.042	&	0.052	&	-	&	-	&	-	&	-	&	-	&	-	&	-	&	-	&	WSRT	\\
8.680	&	-	&	-	&	0.078	&	0.030	&	-	&	-	&	-	&	-	&	-	&	-	&	WSRT	\\
10.258	&	$<$3	&	-	&	-	&	-	&	$<$8.900	&	-	&	-	&	-	&	-	&	-	&	RT-22	\\
\hline
\end{tabular}

\medskip
$^{a}$all measurements in mJy. \\
$^{b}$Telescope key: AMI: AMI Large Array; RT-22: $22$-m RT-22 dish; WSRT: Westerbork Synthesis Radio Telescope; VLA: Very Large Array. \\
$^{c}$Observations from GCN 9533 (Chandra \& Frail 2009).\\
\end{minipage}
\end{table*}

Radio observations were performed with three radio telescopes: the Arcminute Microkelvin Imager (AMI) Large Array, part of the Cavendish Astrophysics Group, Mullard Radio Astronomy Observatory, Cambridge UK; the $22$m RT-22 radio telescope, in Ukraine; and the Westerbork Synthesis Radio Telescope (WSRT), in the Netherlands.  

One epoch was obtained on the AMI on $19$ June $2009$ in the $14.6$ to $17.5$ GHz band at $0.823$ days post burst which led to a positive detection, reported in the GCN Circulars (Pooley 2009a).  The typical noise level of the maps was $0.115$ to $0.190$ mJy, and resolution was typically $40$ x $20$ arcseconds.

Three epochs were obtained on the RT-22 radio telescope: 18 June 2009 ($22$ GHz), 21 June 2009 ($2$ GHz and $8$ GHz) and 28 June 2009 ($2$ GHz and $8$ GHz).  The GRB was not detected in any observation, leading to upper limits only.

Additional radio observations were made with WSRT at $2.3$ GHz and $4.8$ GHz. We used the Multi Frequency Front Ends (Tan 1991) in combination with the IVC+DZB back end\footnote{See sect. 5.2 at http://www.astron.nl/radio-observatory/astronomers/wsrt-guide-observations} in continuum mode, with a bandwidth of 8x20 MHz. Gain and phase calibrations were performed with the calibrators 3C\,48 and 3C~286, at $2.3$ and $4.8$ GHz, respectively.  The observations have been analysed using the Multichannel Image Reconstruction Image Analysis and Display (MIRIAD; Sault et al. 1995) software package. 

The first observation at $4.8$ GHz, $1.7$ days after the burst, was reported in Kamble et al. (2009), but a careful reanalysis of the data resulted in a lower flux value and lower uncertainty in that value.  All of the radio observations are listed in Table \ref{table:090618_radio}.

\subsection{Results \& Discussion}

\subsubsection{The Optical, X-ray \& Radio Afterglow}

\begin{figure}
\centering
\includegraphics[height=3.4in,width=2.5in, angle=270]{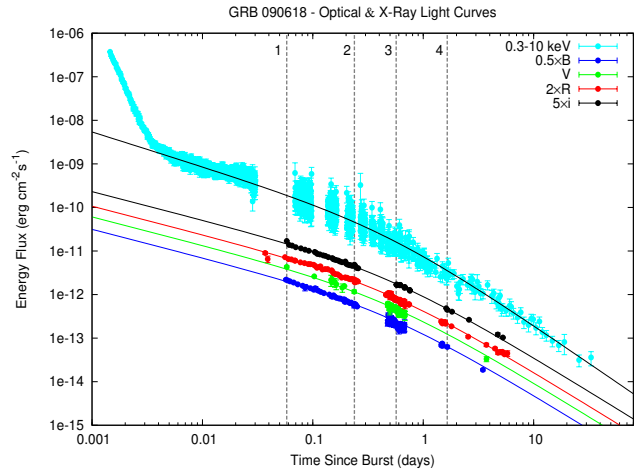}
\caption{GRB 090618: The X-ray and optical light curve.  A broken power-law has been fitted separately to the X-ray ($\chi^{2}/dof=1.17$: $\alpha_{1} = 0.79 \pm 0.01$; $\alpha_{2}= 1.74 \pm 0.04$ and $T_{break} =  0.48 \pm 0.08$ days) and optical data ($\chi^{2}/dof=1.34$: $\alpha_{1} = 0.65 \pm 0.07$; $\alpha_{2} = 1.57 \pm 0.07$ and $T_{break} = 0.50 \pm 0.11$ days).  The dashed grey lines represent the times of the SEDs.  }
\label{fig:090618_OXLC}
\end{figure}

Over 5 hours of optical data were obtained on FTN in $BVR_{c}i$ starting $1.3$ hours after the initial GRB trigger.  This was supplemented by data collected on the other telescopes, resulting in well-sampled light curves.  Figure \ref{fig:090618_OXLC} shows the first six days of optical data: this includes all of the detections in $B$ and $V$, and the first six days of data in $R_{c}$ and $i$.  Several days after the burst the onset of light from an accompanying SN starts to become significant, thus the data after this time cannot be reliably used to characterise \textit{just} the afterglow as it would incorrectly imply a slower afterglow decay.  

Inspection of the panchromatic light curve clearly shows an achromatic break.  Broken power-laws have been simultaneously fit to the $BVR_{c}i$ data from $0.03$ days to $6$ days using Equation (2)\footnote{Similar to our analysis of the afterglow of GRB 060729, we originally fit the optical and X-ray data with the complete Beuermann Function (Beuermann et al. 1999), letting $n$ vary.  Again, the results of our fit found $n \approx 1$, so we fixed $n=1$ when performing the final fit.}, resulting in the following values ($\chi^{2}/dof = 1.38$): $\alpha_{1} = 0.65 \pm 0.08$; $\alpha_{2}= 1.57 \pm 0.07$ and $T_{break} =  0.50 \pm 0.11$ days.  

The 0.3-10 keV light curve, also shown on Figure \ref{fig:090618_OXLC}, has been corrected for foreground and host $N_{H}$ absorption ($N_{H} = 4.0 \pm 0.9 \times 10^{21}$ cm$^{-2}$, see below) and displays the common ``steep-shallow-steep" temporal behaviour (e.g., Nousek et al. 2006; Zhang et al. 2006; Evans et al. 2009) seen for many other GRBs.  The LC is quite featureless (i.e., no flares, etc), which is typical of other classical GRBs with confirmed SNe (e.g., Figure 7 in Starling et al. 2010).  The data at times greater than $t-t_{o} = 0.006$ days were fit with a broken power-law (Equation 2) resulting in the following values ($\chi^{2}/dof = 1.17$): $\alpha_{1} = 0.79 \pm 0.01$; $\alpha_{2}= 1.74 \pm 0.04$ and $T_{break} =  0.48 \pm 0.08$ days.  While the values of $\alpha_{2}$ derived from the optical and X-ray data are relatively similar, they differ by $\approx 2 \sigma$, which could be explained by the presence of a frequency break (i.e., cooling break) between the two wavelength regimes (see section 3.2.3 and Figure \ref{fig:090618_SED4_2}).

We note that the post-break decay index of $\alpha < 2$ is not ``permitted'' in the most simple theories.  However, shallow post-break decays are not a very rare phenomenon, and several examples are seen in the literature (e.g. Zeh et al. 2006).

The column density along the line of sight to the GRB within the host galaxy was estimated from the absorption in the late time X-ray spectrum: the Galactic contribution, $N_{H} = 5.8 \times 10^{20}$ cm$^{-2}$, was fixed to the value measured from $21$cm line radio surveys (Kalberla et al. 2005), while the intrinsic (GRB rest-frame) $N_{H}$ was found to be $N_{H} = 2.7 \pm 0.3 \times 10^{21}$ cm$^{-2}$.  The Galactic and intrinsic $N_{H}$ were obtained with the tbabs and ztbabs models under xspec respectively.  Using $N_{H} = 2.7 \pm 0.3 \times 10^{21}$ cm$^{-2}$ as the rest-frame column density we estimate the amount of extinction expected at optical wavelengths using Equation (4) and Table 2 from Pei (1992) for SMC interstellar dust parameters, finding a modest value of $A_{V} \approx 0.29$ mag.  Similarly low values are also derived for Milky Way (MW) and Large Magellanic Cloud (LMC)-type templates.

\begin{figure}
\centering
\includegraphics[height=3.4in,width=2.5in, angle=270]{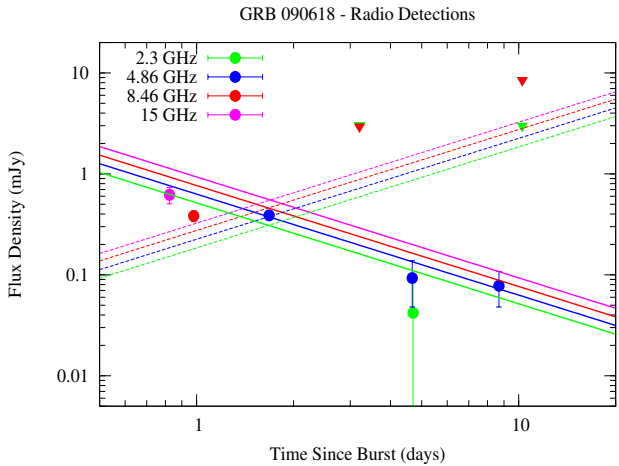}
\caption{GRB 090618: Radio light curves at $2.3$ GHz, $4.8$ GHz, $8.46$ GHz and $16$ GHz, where upper limits are denoted by filled triangles.  Also plotted are the derived radio light curves for a spherical ($f_{\nu}(t) \propto t^{1/2}$) (dotted) and jet-like ($f_{\nu}(t) \propto t^{-1/3}$) (solid) evolution of the afterglow.}
\label{fig:090618_radio}
\end{figure}

The radio detections and upper limits are plotted in Figure \ref{fig:090618_radio}.  Also plotted are our predicted radio LCs (see section 3.2.3).  The radio detections are consistent with a jet-like evolution of the afterglow into an ISM environment.  The modelling of the radio detections will be discussed further in section 3.2.4.

\subsubsection{The Supernova}

\begin{figure}
\centering
\includegraphics[height=3.4in,width=2.5in, angle=270]{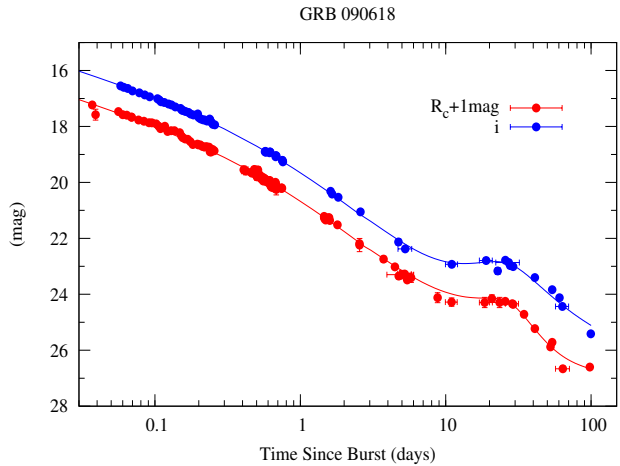}
\caption{GRB 090618: $R_{c}$ (red) and $i$ (blue) light curves of the host-subtracted (see text) magnitudes, which have been corrected for foreground extinction.  The solid line is our model (afterglow and SN1998bw-type SN).  The afterglow is modelled with a broken power-law with parameters: $\alpha_{1} = 0.65 \pm 0.07$; $\alpha_{2} = 1.57 \pm 0.07$ and $T_{break} = 0.50 \pm 0.11$ days, and the template SN (SN1998bw) is dimmed by $0.75$ and $0.5$ mag in $R_{c}$ and $i$ respectively.}
\label{fig:090618_bumps}
\end{figure}

\begin{figure}
\centering
\includegraphics[height=3.4in,width=2.5in, angle=270]{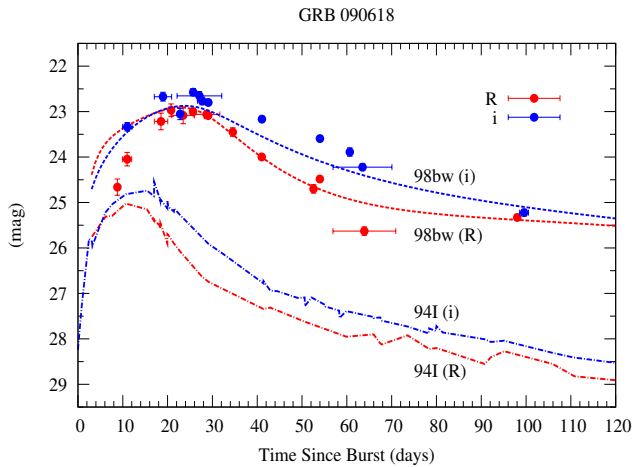}
\caption{GRB 090618: $R_{c}$ (red) and $i$ (blue) supernova light curves of the host and GRB-subtracted (see text) magnitudes.  Plotted for comparison are the light curves of SN1998bw (dashed) and SN1994I (dot-dashed) plotted in the same colours, as they would appear at $z=0.54$.  As seen for the 060729-SN, the brightness and time-evolution of the 090618-SN resembles that of SN1998bw rather than SN1994I.  All magnitudes have been corrected for foreground and host extinction.}
\label{fig:090618_SNe}
\end{figure}

\begin{figure}
\centering
\includegraphics[height=3.4in,width=2.5in, angle=270]{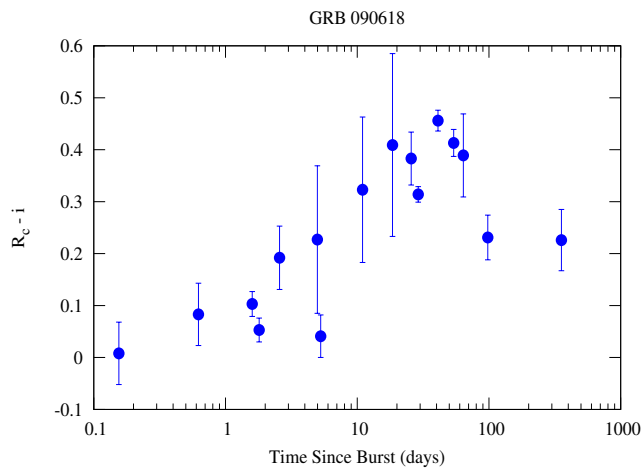}
\caption{GRB 090618: $R_{c}-i$ for the afterglow, SN and host.  The observed change in colour is not constant (as is expected for light from only a GRB) but increases with time.  Such behaviour would be expected if a component of the flux was coming from a core-collapse SN.  At late times the colour is that of the host galaxy ($R_{c}-i \approx 0.22$).}
\label{fig:090618_colour}
\end{figure}

We measured the host magnitudes from our late-time WHT images ($354.7$ days) to be $R_{c} = 23.44 \pm 0.06$ and $i = 23.22 \pm 0.06$.  The flux contribution from the host was mathematically subtracted from all epochs, after all of the magnitudes were converted into fluxes using the zero-points from Fukugita et al. (1995).  The resulting light curves, which have been corrected for foreground extinction, are displayed in Figure \ref{fig:090618_bumps}, and the quoted errors have been calculated in quadrature.  Bumps are clearly detected in both light curves.

We then subtracted the flux due to the GRB from the ``host-subtracted'' light curves using Equation (2), producing the ``SN'' light curves shown in Figure \ref{fig:090618_SNe}.  Also plotted are the distance and extinction (foreground) corrected light curves of 1998bw and 1994I.  As was seen for the SN associated with 060729, the time evolution and brightness of the SN associated with 090618 resembles more closely that of SN1998bw than SN1994I.

We find for the SN associated with GRB 090618 peak apparent magnitudes of $R_{c} = 23.45 \pm 0.14$ and $i = 23.00 \pm 0.09$.  These values, as well as the peak time, are in good agreement with the values predicted by Dado \& Dar (2010). Using the observer-frame $i$-band magnitudes to approximate the rest-frame $V$-band magnitudes, as well as our estimation of the rest-frame extinction ($A_{V}\sim 0.3 \pm 0.1$; see section 3.2.3), we find $M_{V}\ =\ -19.75 \pm 0.13$, which is $\sim 0.3$ mag brighter than SN1998bw in the $V$-band when host extinction is considered (see Table \ref{table:GRB_SNe}).

Additional evidence for an association of a SN with GRB 090618 is the change in $R_{c}-i$ (of the afterglow, SN and host galaxy) over time (Figure \ref{fig:090618_colour}).  At early times a blue colour is seen ($R_{c} - i \approx 0.1$), which we attribute to the afterglow.  However, over time the colour index increases, which is not expected for an afterglow (i.e., solely synchrotron radiation), but is indicative of a component of light coming from a core-collapse supernova. At late times as light from the supernova fades away, the colour index decreases and approaches that of the host galaxy ($R_{c} - i \approx 0.22$, which is typical of GRB host galaxies (e.g., Savaglio et al. 2009: Table 8).

\subsubsection{The Spectral Energy Distribution}

\begin{figure}
\centering
\includegraphics[height=3.4in,width=3.2in, angle=0]{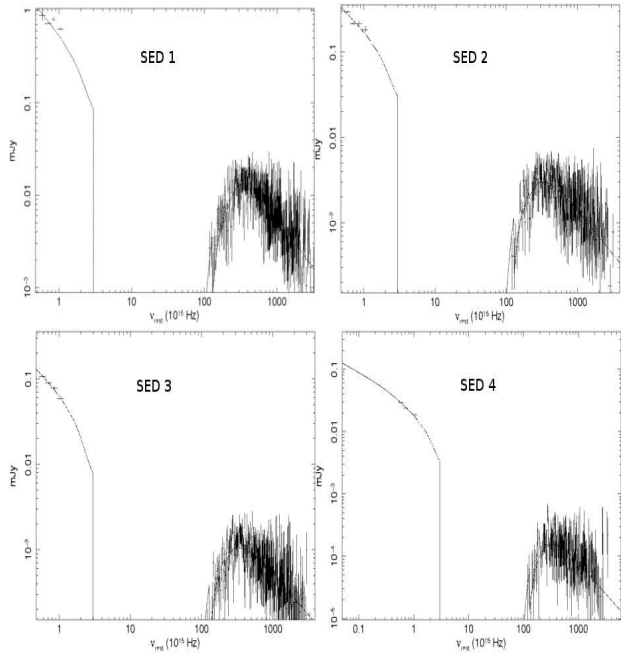}
\caption{GRB 090618: Rest-frame X-ray to optical SEDs: SED1 ($t-t_{o} = 0.059$ days); SED2 ($t-t_{o} = 0.238$ days); SED3: ($t-t_{o} = 0.568$ days); SED4: ($t-t_{o} = 1.68$ days).  The results of the SED modelling are: (1) small rest-frame dust extinction, with $A_{V}\sim 0.3 \pm 0.1$; (2) Each epoch is well fit by a broken power-law (i.e., with a cooling break between the optical and X-ray energy bands) with $\beta_{optical}\sim 0.5$ and $\beta_{xray}= \beta_{optical}+0.5$; (3) the break frequency is clearly decreasing with time, which indicates an ISM environment rather than a wind environment. }
\label{fig:090618_SEDmosaic}
\end{figure}

\begin{table*}
\begin{minipage}{135mm}
\caption{Best-fitting results to SED modelling for GRB 090618  \label{table:090618_SED}}
\begin{tabular}{ccccccc}
\hline
SED & $T-T_{o}$ (days) & $\beta_{optical}$ & $\beta_{xray}$ & $\nu_{c}$ (Hz) & $A_{V,rest}$ (mag) & $\chi^{2}/dof$ \\ 
\hline
SED1 & 0.059 & $0.64^{+0.02}_{-0.01}$ & $1.14^{+0.02}_{-0.01}$ & $3.9^{+0.8}_{-1.2} \times 10^{17}$ & $0.24^{+0.09}_{-0.09}$ & 1.0 \\
SED2 & 0.238 & $0.55^{+0.06}_{-0.07}$ & $1.05^{+0.06}_{-0.07}$ & $5.4^{+4.6}_{-3.5} \times 10^{16}$ & $0.24^{+0.09}_{-0.09}$ & 0.96 \\
SED3 & 0.568 & $0.48^{+0.06}_{-0.06}$ & $0.98^{+0.06}_{-0.06}$ & $1.6^{+2.4}_{-1.0} \times 10^{16}$ & $0.33^{+0.09}_{-0.09}$ & 0.92 \\
SED4 & 1.680 & $0.50^{+0.10}_{-0.11}$ & $1.00^{+0.10}_{-0.11}$ & $7.6^{+6.3}_{-6.2} \times 10^{15}$ & $0.24^{+0.09}_{-0.09}$ & 0.96 \\
\hline
\end{tabular}
\end{minipage}
\end{table*}

\begin{figure}
\centering
\includegraphics[height=3.4in,width=2.5in, angle=270]{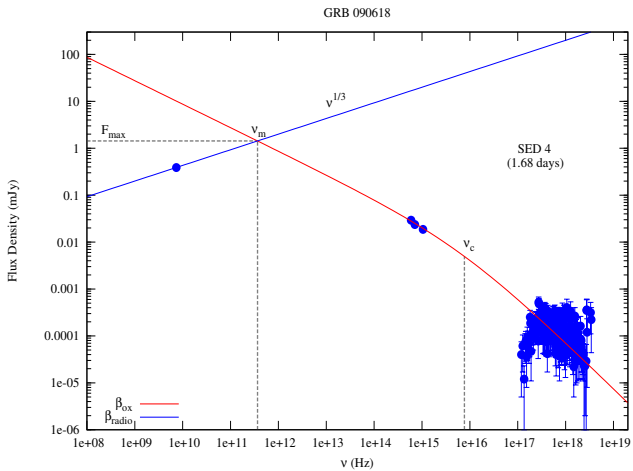}
\caption{GRB 090618: Rest-frame X-ray to radio SED at $t-t_{o} = 1.68$ days.  The typical synchrotron frequency has been estimated from our data, where the optical and X-ray spectral indices have been found from the SED fitting to be: $\beta_{optical} = 0.5$; $\beta_{x} =  \beta_{optical} + 0.5$.  We have used the theoretically-expected value of $\beta_{radio} = 1/3$ and found $\nu_{m} = 3.66 \times 10^{11}$ Hz.  The maximum flux was found to be $F_{max} = 1.43$ mJy. }
\label{fig:090618_SED4_2}
\end{figure}

The rest-frame X-ray to optical SEDs were determined for four epochs: $t-t_{o}=0.059, 0.238, 0.568$ and $1.68$ days.  The times of their occurrence are highlighted on Figure \ref{fig:090618_OXLC} and are referred to in Figure \ref{fig:090618_SEDmosaic} as SED1 to SED4.  For SED1, where there is no contemporaneous X-ray and optical data, the X-ray data have been interpolated to the time of the optical observations.  For all of the SEDs, the X-ray data have been derived from a broader interval so as to collect enough photons, where a normalization coefficient was determined and the interpolation performed through a power-law, which proved to be a good approximation to the data.

The paucity of optical points prevents us from discriminating between the different optical extinction profiles (i.e., MW, LMC \& SMC), thus we considered only the most successful template, i.e., an SMC-type template.  The SEDs were fit using the SMC dust extinction profile from Pei (1992) and combined with a parametrisation of the photometric cross-section to account for the soft X-ray absorption.  They were then applied to a range of models, in particular a simple power-law and a broken power-law.  The simple power-law case was immediately rejected.  Taking the first epoch as an example (similar behaviour was also seen in the other three epochs), the slope of the optical points is $\beta_{opt} \approx 0.3$ while that of the X-rays is $\beta_{X} \approx 1.0$.  These values are not consistent with a common value for $\beta$.  If the optical points had a larger value of $\beta$ than the X-ray, in principle the two values could be reconciled when considering dust effects.  However the opposite case is seen here, indicating the only way to reconcile the two frequency regimes is by invoking a frequency break (i.e., cooling break, $\nu_{c}$) between the two.     

Thus the optical and X-ray spectral indices, as well as the local extinction were allowed to vary during the fit.  After comparison of the initial results it was seen that the results of the fit for each epoch indicated a low value of the extinction, each with a slightly different value.  We thus constrained the extinction within the same range for each epoch (as it is physically unlikely that the extinction would change from epoch to epoch), taking the best determined value from SED3, and allowing it to vary within its range ($0.24<A_{V}<0.42$ mag) and re-fitted the other epochs.  The best-fitting results are displayed in Table \ref{table:090618_SED}.

The temporal evolution of the break frequency was also investigated (Table \ref{table:090618_SED}, Column 5).  Sari, Piran \& Halpern (1999) have shown that the cooling frequency evolves differently depending on the geometry of the afterglow: $\nu_{c} \propto t^{-1/2}$ for a spherical evolution, and $\nu_{c} \approx\ constant$ in a jet-like evolution.  It was seen from the X-ray and optical light curves that a break occurred around $t-t_{o} \approx 0.5$ days, thus SED1 and SED2 occur before the break (i.e., spherical evolution), while SED3 and SED4 occur after the break (i.e., a jet-like evolution).  For the first two epochs, a power-law was fit to the data (i.e., $\nu_{c}=A \times t^{\lambda}$).  The minimum value of the power-law index that can be reasonably fit to the data is $\nu_{c} \propto t^{-0.72 \pm 0.18}$ ($\chi^{2}/dof = 2.08$), which is consistent with that expected from theory.  For the latter two epochs, the large uncertainties derived from the SED fitting (Table \ref{table:090618_SED}) do not allow us to investigate whether the frequency of the cooling break is still evolving or not, thus it can be constant as expected within the derived errors.

The main conclusions we have drawn from the SED modelling are: (1) small rest-frame dust extinction, with $A_{V}\approx 0.3 \pm 0.1$; (2) Each epoch is well fit by a broken power-law (i.e., with a cooling break between the optical and X-ray energy bands) with $\beta_{optical}\sim 0.5$ and $\beta_{xray}= \beta_{optical}+0.5$; (3) the break frequency is clearly decreasing with time, indicating an ISM environment rather than a wind environment (as one expects the break frequency to increase in a wind environment).  The results of the SED fitting appear quite typical of general afterglow broadband spectra, as does its observed time evolution.

\subsubsection{The Jet-break, GRB energetics \& Predicted Radio Light-Curves}

The rest-frame $1-10^{4}$ keV isotropic energy release in gamma-rays was determined by Ghirlanda, Nava \& Ghisellini (2010) in their Table 1: $E_{\gamma,iso} \approx 2.57 \times 10^{53}$ ergs.

A break was seen in the panchromatic LC (Figure \ref{fig:090618_OXLC}), occurring at both optical and X-ray wavelengths at $t-t_{o} \sim 0.5$ days.  If the break is interpreted in the context of a single jet (double sided with sharp edges), using a simple model based upon the standard theoretical framework (e.g., Rhoads 1997; Sari, Piran \& Halpern 1999) and assuming: (1) forward-shock emission; (2) synchrotron radiation; (3) adiabatic evolution of the afterglow; (4) no self-absorption; (5) canonical medium density of $n \sim 1$ cm$^{-3}$; and (6) an isotropic equivalent kinetic energy comparable to that observed in gamma-rays, as well as the definition for $\theta_{jet}$ from Sari, Piran \& Halpern (1999), we estimate an opening angle of $\theta_{jet} \sim 1.5^{o}$.  This in turn implies a corrected, observed gamma-ray emission of $E_{\gamma,\theta} \approx 8.16 \times 10^{49}$ ergs.  If the break is indeed due to a jet a homogeneous ISM is favoured, as a clear break is not expected for a wind environment (e.g., Kumar \& Panaitescu 2000).  

We have been fortunate with our contemporaneous radio detection at $4.86$ GHz at $t-t_{o} = 1.68$ days, and have modelled our radio, optical and X-ray data as the SED shown in Figure \ref{fig:090618_SED4_2}.  We have assumed a jet-like evolution of the afterglow for $t-t_{o} > 0.5$ days (i.e., $f(t) \propto t^{-1.57}$), and neglected effects due to self-absorption, as well as assuming $f_{\nu} \propto \nu^{1/3}$ for frequencies below the typical frequency (i.e., $\nu < \nu_{m}$).  From our SED modelling at $t-t_{o} = 1.68$ days we find $\nu_{sa} \la \nu_{radio} < \nu_{m} < \nu_{optical} < \nu_{c} < \nu_{xray}$  and have estimated the typical synchrotron frequency from our data to be $\nu_{m} = 3.66 \times 10^{11}$ Hz and the peak flux to be $F_{max} = 1.43$ mJy (Figure \ref{fig:090618_SED4_2}).  Using our above assumptions, as well as the derived energetics of the GRB, we have calculated the expected typical frequency of the synchrotron photons and compared it with the observationally-derived value.  In the case where the magnetic field energy density is of order the electron energy density (i.e., $\varepsilon_{B}\approx \varepsilon_{e}$) we estimate $\varepsilon_{B}\approx \varepsilon_{e} \approx 0.045$.  Alternatively, if $\varepsilon_{B} \approx 0.1\varepsilon_{e}$, we estimate $\varepsilon_{B} \approx 0.007$ and $\varepsilon_{e} \approx 0.07$.  Additionally, if we use the value of the cooling/break frequency at $t-t_{o}=1.68$ days ($\nu_{c} = 7.6^{+6.3}_{-6.2} \times 10^{15}$ Hz) that was derived from the SED modelling to estimate $\varepsilon_{B}$ and the medium density $n$, we derive $\varepsilon_{B} \approx 0.001$ if we assume $n\approx 1$.  Alternatively, if we assume $\varepsilon_{B} \approx 0.01$, we derive a medium density $n \approx 0.05$ proton $cm^{-3}$.  These micro-physical values are similar to the assumed values used in the literature (e.g., Zhang et al. 2006).

Using our simple model we have neglected self-absorption effects.  However, this assumption may not be completely reasonable, and the effect of self-absorption would result in a lower typical frequency and a higher peak flux.  We calculate, using typical values of the micro-physical parameters as well as the definition for $\nu_{sa}$ from Granot, Piran \& Sari (1999), a rest-frame self-absorption frequency of $\nu \approx 2.25 \times 10^{10}$ Hz, which is located in the radio frequency range of our detections.  

Assuming that the afterglow has a jet-like evolution after $\approx 0.5$ days, we can estimate the electron energy distribution index $p$ from the optical and X-ray LCs by assuming $t^{-p}$ (i.e., $\alpha = p$) after the break (Sari, Piran \& Halpern 1999).  The optical data imply $p \approx 1.6$ and the X-ray data imply $p \approx 1.75$.  These values are consistent with derived values of $p$ before the break (i.e., for a spherical evolution of the fireball).  Using the X-ray data before the break (i.e., $t^{-3p/4 + 1/2}$ above the cooling frequency), the electron index is found to be $p\approx 1.72$, while from the optical data before the break (i.e., $t^{-3(p-1)/4}$ below the cooling frequency), the value of $\alpha$ implies $p\approx 1.87$.  While these values are lower than the predicted universal value of $p\sim 2.2-2.3$ (Kirk et al 2000; Achterberg et al. 2001), our derived value of $p$ can be accommodated for when it is considered that $p$ cannot have a single value for all GRBs (e.g., Kann et al. 2006; Shen et al. 2006; Starling et al. 2008; Curran et al. 2010), but has a wide range of possible values (e.g., Curran et al. 2010 used X-ray spectral indices of GRB afterglows to parametrize the underlying distribution of $p$, finding the range consistent with a Gaussian distribution centered on $p=2.36$ and having a width of $0.59$.  Our derived value of $p \approx 1.7-1.8$ falls at the lower end of this distribution).

The initial Lorentz factor of the ejecta can be estimated by considering the deceleration time.  Assuming that the onset of the afterglow is masked by the steep decay component (i.e., the tail of the prompt emission), using the definition from Sari \& Piran (1999), and correcting for observer time, we estimate an initial Lorentz factor of $\gamma_{o} > 100$.

The derived radio light curves are shown in Figure \ref{fig:090618_radio}, where the flux scalings have been derived from the X-ray to radio SED at $t-t_{o} = 1.68$ days (Figure \ref{fig:090618_SED4_2}).  Both a spherical ($f_{\nu}(t) \propto t^{1/2}$) and jet-like ($f_{\nu}(t) \propto t^{-1/3}$) evolution of the afterglow are considered, and we conclude that all of the radio observations can be suitably explained by a jet-like evolution of the fireball, though we note that the scarcity of radio detections limits our ability to confidently determine the actual evolution of the afterglow beyond doubt.

\section{Further discussion \& Conclusions}
\subsection{The SNe accompanying GRBs 060729 \& 090618}

We have presented photometric evidence for supernovae associated with GRBs 060729 and 090618.  For GRB 060729, peak apparent magnitudes of the associated supernova are $R_{c} = 23.80 \pm 0.08$ and $I_{c} = 23.20 \pm 0.06$.  When rest-frame extinction was considered using the analysis performed by Schady et al. 2010 ($E(B-V) \le 0.06$ mag), the peak, rest-frame absolute $V$-band magnitude was shown to be $M_{V} = -19.43 \pm 0.06$ which is approximately the same peak brightness as SN1998bw in the $V$-band.

For GRB 090618 the peak apparent magnitudes of the associated supernova are $R_{c} = 23.45 \pm 0.08$ and $I_{c} = 23.00 \pm 0.06$.  The rest-frame extinction local to the GRB was determined from the X-ray to optical SED fitting, and found to be $A_{V} = 0.3 \pm 0.1$.  In turn these values imply a peak, rest-frame absolute $V$-band magnitude of the SN associated with GRB 090618 of $M_{V} = -19.75 \pm 0.13$, which is $\sim 0.3$ mag brighter than SN1998bw in the $V$-band.

\subsection{Comparison with existing GRB \& XRF-associated SNe}

\begin{figure}
\centering
\includegraphics[height=3.4in,width=2.5in, angle=270]{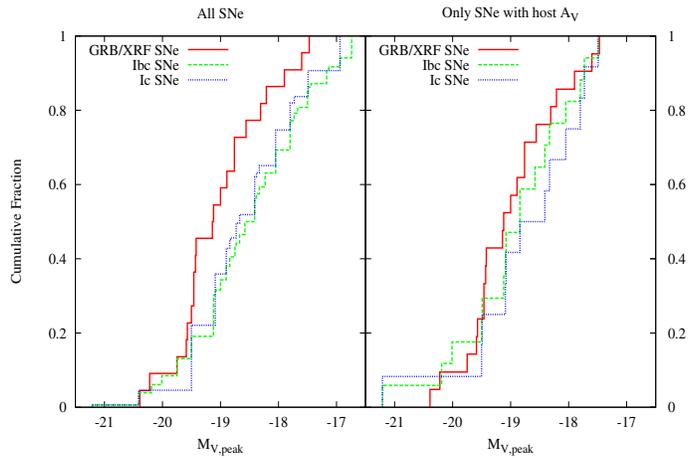}
\caption{Cumulative fraction plots for the two comparison scenarios: ($right$) between \textit{all} SNe and ($left$) considering only those events where an estimation of the host/rest-frame extinction has been made.  In the latter case, the probabilities that the three sets of SNe are drawn from the same parent population are (1) GRB/XRF SNe \& All Ibc SNe: $P=0.88$, (2) GRB/XRF SNe \& only Ic SNe: $P=0.54$, which further supports the idea that GRB-SNe and type Ic SNe have similar progenitors.}
\label{fig:CumulFrac}
\end{figure}

\begin{table*}
\begin{minipage}{155mm}
\caption{Peak Rest-Frame $V$-band Absolute Magnitudes for GRB \& XRF-producing SNe \label{table:GRB_SNe}}
\begin{tabular}{ccccccc}
\hline
GRB & SN & Redshift ($z$) & $A_{V,foreground}$ & $A_{V,host}$ $^{a}$ & $M_{V}^{peak}$ (mag)$^{b,c}$ & Reference \\
\hline
GRB 970228 & - & 0.695 & 0.543 & 0.15  &  $-18.56 \pm 0.30$ &  (1), (2), (3) \\
GRB 980326 & - & $\approx$ 1 & 0.26 & -  & $\approx$ -19.5  & (4) \\ 
GRB 980425 & 1998bw & 0.0085 & 0.18 & 0.05  & $-19.42 \pm 0.30$ & (3), (5), (6), (7), (8), (31)   \\
GRB 990712 & - & 0.434 & 0.09 & 1.67  & $-20.22 \pm 0.20$ & (3), (10), (11), (12), (31) \\
GRB 991208 & - & 0.706 & 0.05 &  0.76 & $-19.46 \pm 0.75$ &  (9), (16) \\
GRB 000911 & - & 1.058 & 0.38 & 0.20  & $-18.31 \pm 0.15$  & (9), (16)  \\
GRB 011121 & 2001ke & 0.36 & 1.33 & 0.39  & $-19.59 \pm 0.33$  & (3), (13), (14), (16) \\
GRB 020405 & - & 0.698 & 0.14 & 0.15  & $-19.46 \pm 0.25$ &  (3), (15), (16), (31)\\
GRB 020410 & - & $\approx\ 0.5$ & 0.40 & 0.0  & $\approx -17.6$ & (3), (17) \\
XRF 020903 & - & 0.251 & 0.09 &  0.0 & $-18.89 \pm 0.30$  &  (3), (18), (31)\\
GRB 021211 & 2002lt & 1.006 & 0.08 & 0.0  & $-18.27 \pm 0.60$ &  (9), (19), (16)\\
GRB 030329 & 2003dh & 0.169 & 0.07 & 0.39  & $-19.14 \pm 0.25$ & (3), (16), (20), (31), (32)  \\
XRF 030723 & - & $\approx\ 0.4$ & 0.089 & 0.23 & $\approx -17.9$ & (3), (9), (34) \\
GRB 031203 & 2003lw & 0.1055 & 2.77 & 0.85  & $-20.39 \pm 0.50$ &  (3), (21), (22), (31)  \\
GRB 040924 & - & 0.859 & 0.18 & 0.16  & $-17.47 \pm 0.48$ & (23) \\
GRB 041006 & - & 0.716 & 0.07 & 0.11  & $-19.57 \pm 0.30$ &  (3), (16), (24)\\
GRB 050525A & 2005nc & 0.606 & 0.25 & 0.32  & $-18.76 \pm 0.28$ &  (3), (25), (26), (33) \\
XRF 060218 & 2006aj & 0.033 & 0.39 & 0.13  & $-18.76 \pm 0.20$ & (3), (27), (28), (31)\\
GRB 060729 & - & 0.54 & 0.11 &  0.18 & $-19.43 \pm 0.06$ & This paper \\
GRB 080319B & - & 0.931 & 0.03 & 0.05  & $-19.12 \pm 0.40$ &  (29), (33) \\
GRB 090618 & - & 0.54 & 0.27 & 0.3  & $-19.75 \pm 0.14$ &  This paper \\
GRB 091127 & 2009nz & 0.49 & 0.12 & 0.0  & $-19.00 \pm 0.20$ & (30) \\
\hline
\end{tabular}

\medskip
$^{a}$Host extinction where available. \\
$^{b}$Cosmological Parameters used: $H_{o}  = 71$ km s$^{-1}$\ Mpc$^{-1}$, $\Omega_{M}  = 0.27$, $\Omega_{\Lambda}  = 0.73$.  \\
$^{c}$Wherever errors are not quoted in the literature conservative errors of $0.4$ mag are used.  \\ \ \\
(1) \cite{Galama00}, (2) \cite{CastLamb99}, (3) \cite{Richardson09}, (4) \cite{Bloom99}, (5) \cite{Galama98}, (6) \cite{McKSch99}, (7) \cite{Sollerman00}, (8) \cite{Nakamura01}, (9) \cite{Zeh04}, (10) \cite{Sahu00}, (11) \cite{Fruchter00}, (12) \cite{Christensen04a}, (13) \cite{Bloom02B}, (14) \cite{Garnavich03}, (15) \cite{Masetti03}, (16) \cite{Kann06}, (17) \cite{Levan05}, (18) \cite{Bersier06}, (19) \cite{DellaValle03}, (20) \cite{Matheson03}, (21) \cite{Malesani04}, (22) \cite{Mazzali06}, (23) \cite{Soderberg06}, (24) \cite{Stanek05}, (25) \cite{DellaValle06a}, (26) \cite{Blustin06}, (27) \cite{Sollerman06}, (28) \cite{Modjaz06}, (29) \cite{Tanvir10}, (30) \cite{Cobb10}, (31) \cite{Levesque10}, (32) \cite{Deng05}, (33) \cite{Kann10}, (34) \cite{Butler05}.
\end{minipage}
\end{table*}

\begin{table*}
\begin{minipage}{140mm}
\caption{Peak Rest-Frame $V$-band Absolute Magnitudes for Local type Ibc \& Ic SNe \label{table:local_SNe}}
\begin{tabular}{ccccccc}
\hline
Type & SN & Redshift ($z$) & $A_{V,foreground}$ & $A_{V,host}$ $^{a}$ & $M_{V}^{peak}$ (mag)$^{b,c}$ & Reference \\
\hline
Ib & 1954A & 0.000977 & 0.07 & - & $-18.75 \pm 0.40$ & (1) \\
Ic & 1962L & 0.00403  & 0.12 & - &  $-18.83 \pm 0.83$ &  (2), (3)\\
Ic & 1964L & 0.002702  & 0.07 & - &  $-18.38 \pm 0.65$ &  (2), (4)\\
Ib & 1966J & 0.002214  & 0.04 & - &  $-19.00 \pm 0.4$ &  (4)\\
Ib & 1972R & 0.002121 & 0.05 & - & $-17.44 \pm 0.4$ & (5) \\
Ic & 1983I & 0.002354  & 0.04 & - &  $-18.73 \pm 0.45$ &  (2), (6)\\
Ib & 1983N & 0.001723 & 0.20 & 0.3 & $-18.58 \pm 0.57$ & (7)\\
Ib & 1983V & 0.005462  & 0.06 & 1.18 &  $-19.12 \pm 0.41$ &  (2), (8)\\
Ib & 1984I & 0.0107 & 0.33 & - & $-17.50 \pm 0.40$ & (9) \\
Ib & 1984L & 0.005281  & 0.08 & 0.0 &  $-18.84 \pm 0.40$ &  (10)\\
Ib & 1985F & 0.00167  & 0.06 & 0.63 &  $-20.19 \pm 0.50$ &  (11)\\
Ic & 1987M &  0.004419 & 0.08 & 1.28 &  $-18.33 \pm 0.71$ &  (2), (12), (13)\\
Ic & 1990B & 0.007518  & 0.10 & 2.53 &  $-19.49 \pm 1.02$ &  (2), (14)\\
Ib & 1991D & 0.041752  & 0.19 & 0.0 &  $-20.01 \pm 0.60$ &  (15)\\
Ic & 1991N & 0.003319  & 0.07 & - &  $-18.67 \pm 1.06$ &  (2)\\
Ic & 1992ar & 0.1451  & 0.30 & 0.0 &  $-18.84 \pm 0.42$ &  (2), (16)\\
Ic & 1994I & 0.001544  & 0.11 & 1.39 &  $-17.49 \pm 0.58$ & (2), (17), (18)\\
Ic BL & 1997ef & 0.011693  & 0.13 & 0.55 &  $-17.80 \pm 0.21$ &  (2), (19), (34)\\
Ic pec & 1999as & 0.127 & 0.09 & 0.0 &  $-21.21 \pm 0.20$ &  (20)\\
Ib/c & 1999cq & 0.026309  & 0.16 & - &  $-19.75 \pm 0.72$ &  (2), (21)\\
Ib & 1999dn & 0.00938 & 0.16 & - &  $-17.17 \pm 0.40$ &  (22)\\
Ib/c & 1999ex & 0.011401  & 0.06 & - &  $-17.67 \pm 0.26$ &  (23)\\
Ib & 2001B & 0.005227 & 0.39 & - & $-17.13 \pm 0.40$ & (24) \\
Ic BL & 2002ap & 0.002187 & 0.29 & 0.0 &  $-17.73 \pm 0.21$ &  (2), (25)\\
Ic & 2003L & 0.021591 & 0.06 & - &  $-18.90 \pm 0.40$ &  (27)\\
Ic BL & 2003jd & 0.018826 & 0.14 & 0.29 &  $-19.50 \pm 0.30$ &  (19), (26), (34) \\
Ic & 2004aw & 0.0175 & 1.15 & 0.0 &  $-18.05 \pm 0.39$ &  (28)\\
Ic & 2004ib & 0.056 & 0.07 & - & $-16.94 \pm 0.40$ & (32) \\
Ib pec & 2005bf & 0.018913 & 0.14 & - & $-18.23 \pm 0.40$ & (33)\\
Ic BL & 2005fk & 0.2643 & 0.19 & - & $-20.41 \pm 0.40$ & (29)\\
Ic BL & 2005kr & 0.13 & 0.31 & 0.27 & $-19.08 \pm 0.40$ & (19), (29), (34)\\
Ic BL & 2005ks & 0.10 & 0.17 & 0.79 & $-18.41 \pm 0.40$ & (19), (29), (34)\\
Ib/c & 2007gr & 0.001728 & 0.19 & - &  $-16.74 \pm 0.40$ &  (30)\\
Ic BL & 2007ru & 0.01546 & 0.89 & 0.0 & $-19.09 \pm 0.20$ & (31)\\
\hline
\end{tabular}

\medskip
$^{a}$Host extinction where available. \\
$^{b}$Cosmological Parameters used: $H_{o}  = 71$ km s$^{-1}$\ Mpc$^{-1}$, $\Omega_{M}  = 0.27$, $\Omega_{\Lambda}  = 0.73$.  \\
$^{c}$Wherever errors are not quoted in the literature conservative errors of $0.4$ mag are used.  \\ \ \\
(1) \cite{Wild60}, (2) \cite{Richardson02}, (3) \cite{Bertola64}, (4) \cite{MillerBranch90}, (5) \cite{Barbon73}, (6) \cite{Tsvetkov83}, (7) \cite{Clocchiatti96}, (8) \cite{Clocchiatti97}, (9) \cite{Binggeli84}, (10) \cite{Tsvetkov87}, (11) \cite{Filippenko86}, (12) \cite{Filippenko90}, (13) \cite{Nomoto90}, (14) \cite{Clocchiatti01}, (15) \cite{Benetti02}, (16) \cite{Clocchiatti00}, (17) \cite{Yokoo94}, (18) \cite{Iwamoto94}, (19) \cite{Modjaz08}, (20) \cite{Hatano01}, (21) \cite{Matheson00}, (22) \cite{Qiu99}, (23) \cite{Martin99}, (24) BAOSS, (25) \cite{Mazzali02}, (26) \cite{Valenti08}, (27) \cite{Soderberg03}, (28) \cite{Taubenberger06}, (29) \cite{Baretine05}, (30) \cite{Foley07}, (31) \cite{Sahu09}, (32) \cite{Adelman05}, (33) \cite{Anupama05}, (34) \cite{Levesque10}.
\end{minipage}
\end{table*}

\begin{table*}
\begin{minipage}{165mm}
\caption{Kolmogorov-Smirnov test results \label{Table:KS}}
\begin{tabular}{ccccccc}
\hline
Dataset  & Number of Data points & Mean & Standard Deviation & $P^{a}$ & $D^{a}$ & comments \\
\hline
GRB/XRF-associated SNe & 22 & -19.02 & 0.77  & - & - & all events \\
Local type Ic SNe & 19 & -18.73 & 1.00  & 0.16 & 0.33 & all events \\
Local type Ibc SNe & 34 & -18.59 & 1.04  & 0.12 & 0.31 & all events \\
\hline
GRB/XRF-associated SNe & 21 & -19.00 & 0.78  & - & - & only those with host $A_{V}$ \\
Local type Ic SNe & 12 & -18.75 & 1.03 & 0.54 & 0.27 & only those with host $A_{V}$  \\
Local type Ibc SNe & 17 & -18.93 & 0.97  & 0.88 & 0.18 & only those with host $A_{V}$  \\
\hline
\end{tabular}

\medskip
$^{a}$Probability and maximum difference between the GRB/XRF SNe sample and the local SNe sample.
\end{minipage}
\end{table*}

To put our detections into context, we compiled from the literature peak, rest-frame $V$-band absolute magnitudes for two samples of type Ibc SNe, incorporating the values of the host/rest-frame extinction (taken at the location of the GRB or SN when available; i.e., Kann et al. 2006; Modjaz et al. 2008; Levesque et al. 2010), and applying the SMC reddening law.  The two samples are: (1) those associated with GRBs \& XRFs, and (2) local type Ibc SNe.  We note that we have limited this analysis to incorporate only GRB/XRF events where an optically-bright SN has been positively detected.  These samples, along with their respective references, are listed in Tables \ref{table:GRB_SNe} and \ref{table:local_SNe}.  

When comparing these two samples we are attempting to answer the question: ``Are the progenitors of GRB \& XRF associated SNe the same as those of local type Ibc SNe without an accompanying GRB/XRF trigger?''  by testing if the distribution of the peak magnitudes of the two samples of supernovae are different.  To do this we performed a Kolmogorov-Smirnov (KS) test on the two samples. First we compared the entire GRB/XRF sample ($N=22$) with \textit{all} of the type Ibc SNe ($N=34$), finding a modest probability that the two samples are drawn from the same parent population of $P=0.12$.  When we compare the GRB/XRF sample with only the type Ic SNe ($N=19$) we find a similar probability of $P=0.16$.  

However, when we limited the samples to include only those SNe where an estimation of the host/rest-frame extinction has been made, we see an increased probability between the datasets: $P=0.88$ between the GRB/XRF SNe ($N=21$) and all of the type Ibc SNe ($N=17$), and $P=0.54$ between the GRB/XRF SNe and only the type Ic SNe ($N=12$).  For these samples a higher average peak magnitude is also seen among the local type Ibc SNe sample, with the local type Ibc SNe having $M_{V,Ibc} = -18.93$.  This is an increase in average peak brightness of $\sim 0.3$ mag, and is likely due to the inclusion of five additional bright type Ib events with known host-extinction (SN: 1983N, 1983V, 1984L, 1985F \& 1991D) with the local Ic SNe sample (N=12).  The average peak magnitude of these five SNe is $M_{V,Ib} = -19.34$, which places them among the brightest of the type Ib events.  For comparison, a study by Richardson et al. (2002) analysed the absolute peak magnitudes of samples of all types of nearby SNe, and distinguished between ``Normal'' and ``Bright'' type Ibc events, with the latter having $M_{V,Ibc} = -19.72 \pm 0.24$, comparable with the average peak magnitudes of these five SNe.  Thus the inclusion of these 5 bright Ib events has increased the average peak brightness of the type Ibc SNe sample, as well as increased the probability of association between our samples of host extinction-corrected type Ibc SNe and the GRB/XRF-SNe.  The results of the K-S test are summarized in Table \ref{Table:KS} and cumulative fraction plots are shown in Figure \ref{fig:CumulFrac}.

It is worth noting, however, that while the increase in probability may really be due to an increased association between the datasets, the caveat of smaller sample sizes is that as one goes to smaller samples it is harder to obtain a statistically significant discrepancy.  

Our analysis is not unique, with previous studies having performed similar analysis, which our results generally support.  Richardson 2009 (R09) found from a sample of 14 GRB/XRF-SNe with (mostly) known values of the host extinction an average $M_{V,peak} = -19.2 \pm 0.2$, with $\sigma =0.7$, which our results are in agreement with.  R09 also compared the GRB/XRF-SNe sample with a sample of stripped-envelope SNe, which included types Ib, Ic and IIb finding the GRB/XRF-SNe sample brighter by $\sim 0.8$ mag.  We find that our complete GRB/XRF-SNe sample is $\sim 0.4$ mag brighter than the complete local type Ibc SNe (but only $\sim 0.1$ mag when only events with known host extinction are considered), which is somewhat less than R09, however our sample does not include any type IIb SNe.

Ferrero et al. 2006 (F06) undertook a slightly different study, comparing the existing GRB/XRF-SNe sample (to date) with that of XRF 060218.  They furthered an original analysis performed by Zeh et al. (2004), and calculated for each GRB-SN event the luminosity ratio $k$ and stretch factor $s$ in comparison with SN1998bw.  They find for their host extinction-corrected sample of GRB/XRF-SNe clustering in the range $0.6 <\ k\ <\ 1.5$, implying a range of absolute magnitudes at peak brightness of $+0.55$ mag $ < \ M_{V}^{98bw}\ <\ -0.44$ mag, which is in good agreement with our results.  They conclude that the width of the GRB/XRF-SN luminosity function is at least 2 mag, and there was no evidence that the luminosity function evolved with redshift.  Our analysis concurs with both of these results.

In conclusion, while the complete sample of GRB/XRF SNe are generally brighter ($M_{V,GRB} = -19.02$, $\sigma = 0.77$) than the complete local type Ibc SNe ($M_{V,Ibc} = -18.59$, $\sigma = 1.04$) and the type Ic SNe sample ($M_{V,Ic} = -18.73$, $\sigma = 1.00$), our test does not address factors such as host and progenitor metallicity and typical outflow velocities.  Thus we cannot rule out the null-hypothesis that the samples of SNe are drawn from the same parent population.

\ \\
\indent We are very grateful to Knut Olsen and Abi Saha for their assistance in obtaining images on the CTIO $4$-m telescope in August 2006.  We would also like to thank the anonymous referee for their very thorough and constructive comments of the original manuscript.  This work was supported partially by a Science and Technology Facilities Council (STFC) (UK) research studentship (ZC).  MI, YJ, and WKP were supported by the Korea Science and Engineering Foundation (KOSEF) grant No. 2009-0063616, funded by the Korea government (MEST).  The T40 telescope at the Observatorio de Aras de los Olmos is funded by the Generalitat Valenciana, the Spanish Ministerio de Ciencia e Innovacion and the Universitat de Valencia.  The research of JG and AJCT is supported by the Spanish programmes ESP2005-07714-C03-03, AYA2007-63677, AYA2008-03467/ESP and AYA2009-14000-C03-01. TAF and ASM were supported by the grant of the President of the Russian Federation (MK-405.2010.2).  The WSRT is operated by ASTRON with financial support from the Netherlands Organization for Scientific Research (NWO).  AG acknowledges funding from the Slovenian Research Agency and from the Centre of Excellence for Space Sciences and Technologies SPACE-SI, an operation partly financed by the European Union, European Regional Development Fund and  Republic of Slovenia, Ministry of Higher Education, Science and Technology.  A.F.S. acknowledges support from the SpanishMICINN projects AYA2006-14056, Consolider-Ingenio 2007-32022, and from the Generalitat Valenciana project Prometeo 2008/132.

{}

\end{document}